\documentclass[preprint]{iacrtrans}

\usepackage{epsfig}
\usepackage{amssymb}
\usepackage{amsmath}
\usepackage{multicol}
\usepackage{colortbl}
\usepackage{booktabs}
\usepackage{multirow}
\usepackage[subtle]{savetrees}
\usepackage[utf8]{inputenc}

\usepackage{svg}

\usepackage{amsfonts}
\usepackage{graphicx}
\usepackage{wrapfig}
\graphicspath{ {images/} }
\usepackage{algorithmic}
\usepackage[boxruled,vlined,linesnumbered]{algorithm2e}
\usepackage[n ,advantage ,operators ,sets, adversary,landau,probability,notions,logic,ff,mm,primitives,events,complexity,asymptotics,keys]{cryptocode}
\usepackage{caption}
\usepackage{float}
\usetikzlibrary{matrix,shapes,arrows,positioning,chains, calc}

\usepackage{tikz}
\usetikzlibrary{fit}
\usetikzlibrary{shapes,arrows,positioning}

\usepackage{amsmath}
\usepackage{amsthm} 
\usepackage{hyperref}
\usepackage{centernot}

\usepackage{lscape}

\usepackage{enumitem}
\usepackage{bigdelim}
\usepackage{adjustbox}
\usepackage{subcaption}
\usepackage{bm}
\usepackage{url}
\usepackage{breakcites}

\restylefloat*{figure}
\restylefloat*{table}

\usepackage[]{todonotes}

\SetKwData{Left}{left}\SetKwData{This}{this}\SetKwData{Up}{up}
\SetKwFunction{Union}{Union}\SetKwFunction{FindCompress}{FindCompress}
\SetKwInOut{Input}{Input}\SetKwInOut{Output}{Output}

\author{Suparna Kundu\inst{1}, Archisman Ghosh \inst{2}, Angshuman Karmakar\inst{3}, \\Shreyas Sen\inst{2} \and Ingrid Verbauwhede\inst{1}}
\institute{
  COSIC, KU Leuven, Belgium
  \and
  Purdue University, USA
  \and
  Indian Institute of Technology Kanpur, India\\
  \email{suparna.kundu@esat.kuleuven.be},\ 
  \email{ghosh69@purdue.edu},\ 
  \email{angshuman@cse.iitk.ac.in},\ 
  \email{shreyas@purdue.edu},\ 
  \email{ingrid.verbauwhede@esat.kuleuven.be}
}
\authorrunning{Kundu et al.}

\title[Rudraksh]{Rudraksh: A compact and lightweight post-quantum key-encapsulation mechanism}

\begin{document}

\maketitle

\keywords{Post-quantum cryptography, key-encapsulation mechanism, Lightweight cryptography, Lattice-based cryptography, Hardware implementation, FPGA.}


\begin{abstract}
Resource-constrained devices such as wireless sensors and Internet of Things (IoT) devices have become ubiquitous in our digital ecosystem. These devices generate and handle a major part of our digital data. However, due to the impending threat of quantum computers on our existing public-key cryptographic schemes and the limited resources available on IoT devices, it is important to design lightweight post-quantum cryptographic (PQC) schemes suitable for these devices.

In this work, we explored the design space of learning with error-based PQC schemes to design a lightweight key-encapsulation mechanism (KEM) suitable for resource-constrained devices. We have done a scrupulous and extensive analysis and evaluation of different design elements, such as polynomial size, field modulus structure, reduction algorithm, {and} secret and error distribution of an LWE-based KEM. Our explorations led to the proposal of a lightweight PQC-KEM, Rudraksh, without compromising security. Our scheme provides security against chosen ciphertext attacks (CCA) with more than 100 bits of Core-SVP post-quantum security and belongs to the NIST-level-I security category (provide security at least as much as AES-128). We have also shown how ASCON can be used for lightweight pseudo-random number generation and hash function in the lattice-based KEMs instead of the widely used Keccak for lightweight design. Our FPGA results show that Rudraksh currently requires the least area among the PQC KEMs of similar security. Our implementation of Rudraksh provides a $\sim3\times$  improvement in terms of the area requirement compared to the state-of-the-art area-optimized implementation of Kyber, can operate at $63\%$-$76\%$ higher frequency with respect to high-throughput Kyber, and improves time-area-product $\sim2\times$ compared to the state-of-the-art compact implementation of Kyber published in HPEC 2022. 

\end{abstract}

\section{Introduction}\label{sec:intro}

Lightweight cryptography (LWC) is a niche research area in cryptography that studies methods to incorporate secure cryptographic protocols into devices with minimal resources due to their operational requirements. 
%
%
There are two major avenues in the research and development of LWC. 
First, implement existing cryptographic protocols that are not specifically designed as LWC in {a} \emph{lightweight manner} such as lightweight implementations of symmetric-key ciphers such as AES (Advanced Encryption Standard)~\cite{AES_compact_1, AES_compact_2}, Keccak~\cite{keccak_compact_1, keccak_compact_2}, or public-key cryptographic (PKC) algorithms such as RSA~\cite{RSA} and elliptic curve cryptography (ECC)~\cite{ECC_compact_1, ECC_compact_2}.
Second, design cryptographic schemes that are lightweight implementation \emph{friendly} such as symmetric-key cipher ASCON~\cite{ASCON}, which is the winner of the National Institute of Standards and {Technology's (NIST's)} lightweight cryptography competition~\cite{NIST_LWC} and also selected as the `primary choice' for lightweight authenticated encryption in the final portfolio of the CAESAR~\cite{caeser} competition. Another example is Quark~\cite{quark}, which is a lightweight hash function designed specifically for low-power devices such as Radio Frequency Identification (RFID) devices. 

{Recently, NIST also standardized post-quantum cryptography (PQC) schemes in anticipation of the arrival of large-scale quantum computers and their detrimental effect on our existing PKC schemes. These are key-encapsulation mechanism (KEM) {CRYSTALS}-Kyber~\cite{kyber_specification} and digital signature schemes SPHINCS+~\cite{web:sphincs}, {CRYSTALS}-Dilithium \cite{dilithium}, and Falcon~\cite{web:falcon}}{.}
Naturally, we will also have to equip resource-constrained Internet of Things (IoT) and embedded devices with quantum-secure cryptographic schemes to secure them for the foreseeable future. {Although it should be noted that there exist some LW implementations of the standardized schemes such as the compact implementation Kyber~\cite{9926344} or Dilithium~\cite{compact_dilithium_1, compact_dilithium_2}.}

\begin{figure}[!ht]
  \centering
  \includegraphics[scale=.5]{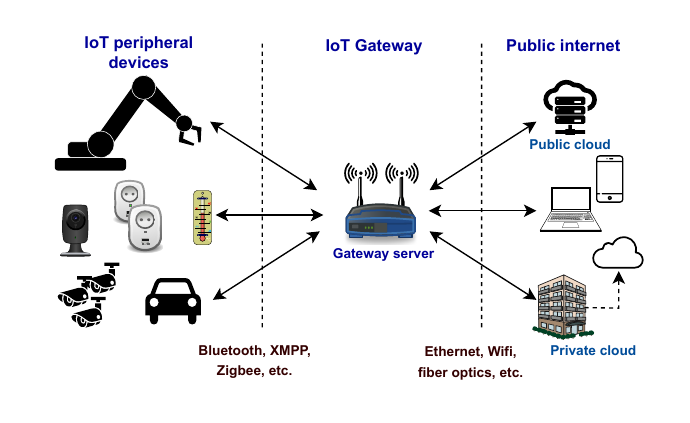}
  \vspace{-10pt}
  \caption{An illustrative example of a typical IoT gateway architecture.}
  \label{fig:iot_gateway}
\end{figure}

Apart from LW design, there is another subtle issue in incorporating cryptographic schemes into IoT devices. Consider a typical IoT ecosystem, as shown in Fig.~\ref{fig:iot_gateway}. Here, the IoT peripheral devices connect to the public internet through an IoT gateway server. The IoT gateway architecture has several layers, such as a security layer, device layer, data management layer, etc. As part of connecting IoT peripheral devices, the IoT gateway servers perform data filtering and processing, protocol translation, authorization and authentication, etc.  Whenever a user or device wants to connect to a peripheral device or vice-versa, the gateway servers have to run proper authentication and authorization protocols to make this happen. The gateway servers are usually powerful servers serving numerous IoT peripheral devices simultaneously. So, for them, high throughput is a more important operational metric than resource consumption. Meanwhile, the reverse is more important for IoT peripheral devices, which connect sporadically to the IoT gateway servers. Therefore, a \textit{flexible} cryptographic scheme that can be instantiated either in a high latency and low resource consumption mode or in a low latency and high resource consumption mode is highly suitable in this scenario.

{From a very broad perspective, this work aims to push the lower bound of the resource consumption of post-quantum cryptography, especially learning with errors (LWE)-based post-quantum cryptography.}
For the rest of this work, we will use the term lattice-based cryptography (LBC) to denote the cryptographic schemes based on the learning with errors (LWE)~\cite{DBLP:journals/jacm/Regev09} problem or its variants such as ring-learning with errors~\cite{DBLP:conf/eurocrypt/LyubashevskyPR10}, learning with rounding~\cite{Banerjee2011}, etc. We also want to delineate the term \emph{lightweight} implementation here. For software implementation on resource constraint devices like Cortex-M0/M4 we use the term lightweight implementation for implementations with low memory footprint such as~\cite{low_memory_kyber,saber_on_arm,compact_dilithium_low_memory}. Lightweight design also implies low area and low-power or energy solutions for hardware devices such as field-programmable gate arrays (FPGA) or application-specific integrated circuits (ASIC); however, low-power or energy implementation cannot be demonstrated without custom ASIC design as FPGA defaults to high power-consuming interfaces. We expect that the ASIC version of our design will also reflect relatively lower memory as we can custom-make memory as per our requirement instead of using entire Block RAMs. We have demonstrated a low-area implementation on FPGA in this work. We have kept ASIC-related optimization as part of future work. Below, we briefly summarize the salient contributions of this work.

\noindent
\textbf{PQ KEM suitable for resource-constrained devices: }We propose a lightweight post-quantum chosen-ciphertext attack (CCA) secure module-learning with errors-based key-encapsulation mechanism (MLWE-KEM), {Rudraksh}. 
{In {1998},~\cite{NTRU} proposed NTRUEncrypt, a public-key encryption scheme that can achieve high speed and require low memory. Years later,~\cite{DBLP:conf/ccs/BuchmannGGOP16} deliberated on a lightweight CPA-secure LWE-based key-exchange scheme suitable for IoT devices before us. However, none of these schemes are secure under chosen ciphertext attacks. Hermans et al.~\cite{hermans_iot} have pointed out that a CPA-secure scheme can only provide security against a narrow range of adversaries. Therefore, in this work, we focus on designing a CCA-secure KEM. Kindly note that several CCA-secure NTRU-based KEM have been proposed during the NIST PQC standardization procedure. Most lattice-based KEMs, such as Kyber, Saber~\cite{saber_specification_3rd_round} (LightSaber), NewHope~\cite{newhope}, and NTRU-based KEMs~\cite{ntru_kem_nist_submission}, have a low-security version design. Still, these are not explicitly designed to be lightweight, specifically for resource-constrained devices.}
One of the most prominent issues among the standardized PQC signatures is their large signature size compared to {the} classical signature schemes. This has a very detrimental effect on some protocols, such as transport layer security (TLS), where the increase in the size of certificates in the chain-of-trust model leads to serious performance degradation~\cite{sikeridis_TCP_congestion} due to the congestion control mechanism of the transmission control protocol (TCP). To address such problems, NIST has called for another standardization~\cite{nistadditional} for PQ signatures with small signature sizes and fast verification time. {In addition}, some proposals have been made {to replace} the TLS handshake with a CCA-secure KEM, such as KEMTLS~\cite{kemtls_schwabe} or TLS-PDK~\cite{KEMTLS_PDK_Schwabe}, for better performance. Therefore, we believe that lightweight KEMs {can} profoundly impact the transition from classical PKC to PQC. Further, the techniques developed in this work can also be used to create a lightweight PQ digital signature scheme.

\noindent
\textbf{Practical design strategy: }We adopt a new design style that is strongly coupled with hardware implementation. 
Our design decisions have been strongly driven by their potential advantage for lightweight hardware implementation. We explored the parameter space of {LWE-based KEMs} to propose optimum parameters that satisfy our design objectives. The NIST PQ standardization procedure witnessed a collective effort from researchers around the world for a thorough and rigorous analysis of different design elements of LBC. Therefore, to reap the benefits of the procedure and to bolster confidence in our designed KEM {Rudraksh}, we have kept the design very Kyber-\textit{esque}. 
{We refrained from making aggressive design decisions such as using non-constant-time modular reductions. We explore different centered binomial distributions with different variations for sampling error and secret vectors. However, we have not explored other distributions, such as the binary distribution~\cite{DBLP:conf/ccs/BuchmannGGOP16} or the fixed
weight distribution~\cite{round5, smaug_kem, TiGER}.}
We also explored and analyzed different implementations of number theoretic transform (NTT)-based polynomial multiplication. We have used a modular reduction algorithm for lightweight hardware implementation. {We also provided an optimized hardware implementation using the Xilinx Virtex-7 and Artix-7 FPGA to demonstrate the efficacy and justify our design decisions.} We have also extensively compared our scheme with other state-of-the-art compact implementations of LBC.

\noindent
\textbf{Lightweight implementation: }We demonstrated our design of {Rudraksh} using a lightweight implementation on FPGA. We observe that a major source of hardware overhead for lattice-based KEMs such as Kyber comes {mainly} from the bulky Keccak~\cite{keccak_ec} module, storing the twiddle factors for NTT, and other memory requirements. We want to make this KEM design suitable for lightweight CCA-secure KEMs, so we focus on minimizing the area in terms of FPGA resources such as LUT, DSP, flip-flops, and memory with reasonable latency {(54-81 $\mu$s)}. ASCON~\cite{ASCON} is a family of lightweight authenticated encryption and hashing algorithms. 
{We have replaced Keccak with ASCON~\cite{ASCON} to reduce the overhead of Keccak. This also demonstrates the efficiency and benefits of using ASCON in a {LBC} scheme.}
{Our results show that area and memory can be reduced by approximately $\sim3\times$  with respect to the most resource-optimized Kyber~\cite{HE2024167}. Our pipelined hardware implementation of Rudraksh can operate at 63\%-76\% higher frequency compared to high-throughput Kyber~\cite{DBLP:journals/iacr/DangFAMNG20} and also provides $\sim2\times$  time-area-product improvement compared to the compact implementation of Kyber~\cite{9926344}.}

\vspace{-10pt}
\section{Background and related works}\label{sec:prelims}
We denote the set of integers modulo $q$ as $\mathbb{Z}_q$ and the quotient ring $\mathbb{Z}_q/(x^n+1)$ as {$R_q$} ($n\geq1$). 
The ring containing the vectors with {$\ell$} elements and the matrix with {$\ell \times \ell$} elements from {$R_q$} are represented as {$R_q^{(\ell)}$ and $R_q^{(\ell \times \ell)}$,} respectively.
Lowercase letters indicate polynomials ($v\in${$R_q$}), and bold lowercase letters denote vectors of polynomials ($\pmb{s}\in${$R_q^{(\ell)}$}). Bold uppercase letters represent matrices of polynomials ($\pmb{A} \in${$R_q^{(\ell \times \ell)}$}). Multiplication of two polynomials $a\in${$R_q$} and $b\in${$R_q$} is denoted by $a\cdot b\in${$R_q$}. The number theoretic transform (NTT) representation of a polynomial $a \in${$R_q$} is denoted by $\hat{a}$. 
{The point-wise multiplication between these two polynomials in the NTT domain $\hat{a}$ and $\hat{b}$ is presented by $(\hat{a}\circ \hat{b})$.} 
When NTT is applied to each constituent element of $\pmb{a} \in${$R_q^{(\ell)}$} and $\pmb{A} \in${$R_q^{(\ell \times \ell)}$}, it is denoted as $\pmb{\hat{a}}$ and $\pmb{\hat{A}}$ respectively. 
{$\pmb{\hat{a}}\circ \pmb{\hat{b}}$ represents vector-vector point-wise multiplication between vector of polynomials $\pmb{\hat{a}}$ and vector of polynomials $\pmb{\hat{b}}$. $\pmb{\hat{A}}\circ \pmb{\hat{b}}$ denotes matrix-vector point-wise multiplication (PWM) between {a} matrix of polynomials $\pmb{\hat{A}}$ and {a} vector of polynomials $\pmb{\hat{b}}$.} 
$a\leftarrow\chi(S)$ represents that $a$ is sampled from the set $S$ according to the distribution $\chi$, and we use $\leftarrow$ to denote probabilistic output. $a:=\chi(S;\ {\mathrm{seed}}_a)$ indicates that $a\in S$ is generated from the ${\mathrm{seed}}_a$ and follows the distribution $\chi$, and we use $:=$ to denote deterministic output. We use $\mathcal{U}$ to denote uniform distribution and $\beta_{\mu}$ to denote the centered binomial distribution (CBD) with standard deviation $\sqrt{\mu/2}$. We denote the {Hamming} weight function by $\mathtt{HW}$. $|x|$ denotes bit length of the bitstring $x$.

\vspace{-5pt}
\subsection{Learning with Errors Problem} 
The learning with errors (LWE) problem was introduced by Regev~\cite{DBLP:journals/jacm/Regev09} and is as hard as standard worst-case lattice problems~\cite{DBLP:conf/stoc/Peikert09}. Given $\pmb{A} \leftarrow \mathcal{U}(\mathbb{Z}_q^{(m\times n)})$, $m=O({\mathrm{poly}}(n))$ $\pmb{s} \leftarrow \chi_1(\mathbb{Z}_q^{(n)})$, and $\pmb{e} \leftarrow \chi_2(\mathbb{Z}_q^{(m)})$, where $\chi_1$ and $\chi_2$ are two narrow distributions. The LWE instance consists of the pair $(\pmb{A},\ \pmb{A}\cdot\pmb{s}+\pmb{e}) \in \mathbb{Z}_q^{(m\times n)}\times \mathbb{Z}_q^{(m)}$. The LWE problem states that for $b\leftarrow\mathcal{U}(\mathbb{Z}_q^{(m)})$, it is hard to distinguish between the pairs $(\pmb{A},\ \pmb{A}\cdot\pmb{s}+\pmb{e})$ and $(\pmb{A},\ \pmb{b})$. The hardness depends on the distributions $\chi_1, \chi_2$ and the parameters $q,\ n$.
The Ring-LWE (RLWE)~\cite{DBLP:conf/eurocrypt/LyubashevskyPR10} and the Module-LWE (MLWE)~\cite{module_sis} problems are algebraically structured variants of the LWE problem. $\pmb{A},\ \pmb{s},\ \pmb{e}$ are polynomials sampled from the ring ${R_q}=\mathbb{Z}_q/(x^n+1)$ in the RLWE problem.
In the MLWE problem, $\pmb{A}$ is a matrix of polynomials sampled uniformly from {$R_q^{(\ell \times \ell)}$}, and $\pmb{s},\ \pmb{e}$ are vectors of polynomials {sampled} from the set {$R_q^{(\ell)}$}. 

\vspace{-5pt}
\subsection{MLWE-based Public-key Encryption}\label{lwe-pke}
\begin{figure}[!t]
\vspace{-10pt}
\centering
\scriptsize
\fbox{\begin{varwidth}{\textwidth}
\begin{subfigure}[t]{0.47\textwidth}
    \raggedright 
    \begin{small}
    $\mathtt{PKE}{.}\mathtt{KeyGen} ()$
    \begin{enumerate}[wide=0em, itemsep=0pt, parsep=0pt, font=\scriptsize\tt\color{gray}]
            \item $\textcolor{black}{\mathrm{seed}}_{\pmb{A}},\ \textcolor{black}{\mathrm{seed}_{\pmb{\mathrm{se}}}} \leftarrow  \mathcal{U}(\{0,\ 1\}^{\text{len}_{K}}\textcolor{black}{\times \{0,\ 1\}^{\text{len}_{K}}})$ \\
            \item $\pmb{\hat{A}} :=  \mathtt{PRF}(\textcolor{black}{R_q^{(\ell \times \ell)}};\  \textcolor{black}{\mathrm{seed}}_{\pmb{A}})$ \ \ \ 
  $\triangleright\ (\pmb{\hat{A}} = \mathtt{NTT} (\pmb{A}))$ \\
            \item \textcolor{black}{$\pmb{s},\ \pmb{e} :=  \beta_{\eta} ( \textcolor{black}{R_q^{(\ell)}} \textcolor{black}{\times R_q^{(\ell)}} ;\  \textcolor{black}{\mathrm{seed}}_{\pmb{\mathrm{se}}})$} \\
            \item $\pmb{\hat{s}} := \mathtt{NTT} (\pmb{s}) \in \textcolor{black}{R_q^{(\ell)}},\ \pmb{\hat{e}} := \mathtt{NTT} (\pmb{e}) \in \textcolor{black}{R_q^{(\ell)}}$ \\
            \item $\pmb{\hat{b}} :=  (\pmb{\hat{A}}\circ \pmb{\hat{s}} + \pmb{\hat{e}}) \in \textcolor{black}{R_q^{(\ell)}}$ \\
            \item \textbf{return} $(\textcolor{black}{\mathrm{pk}} := (\textcolor{black}{\mathrm{seed}}_{\pmb{A}},\  \pmb{\hat{b}}),\  \textcolor{black}{\mathrm{sk}} :=  (\pmb{\hat{s}}))$
    \end{enumerate} 
    \end{small}
\end{subfigure}
\begin{subfigure}[t]{0.50\textwidth}
    \raggedright
    \begin{small}
    $\mathtt{PKE}{.}\mathtt{Enc} (\textcolor{black}{\mathrm{pk}} := (\textcolor{black}{\mathrm{seed}}_{\pmb{A}},\  \textcolor{black}{\pmb{\hat{b}}}),\  m \in \textcolor{black}{R_{2^B}};\  r )$
    \begin{enumerate}[wide=0em, itemsep=0pt, parsep=0pt, font=\scriptsize\tt\color{gray}]
        \item $\pmb{\hat{A}} \leftarrow \mathtt{PRF}(\textcolor{black}{R_q^{(\ell \times \ell)}};\  \textcolor{black}{\mathrm{seed}}_{\pmb{A}}) $ \\
        \item \textbf{if } $r$  is not specified \textbf{then} $ r \leftarrow  \mathcal{U}(\{0,\ 1\}^{256})$\\
        \item \textcolor{black}{$\pmb{s'},\ \pmb{e'} :=  \beta_{\eta} ( \textcolor{black}{R_q^{(\ell)}}\textcolor{black}{\times R_q^{(\ell)}} ;\  r)$} \\
        \item \textcolor{black}{${e''} :=   \beta_{\eta} ( \textcolor{black}{R_q} ;\  r||2\ell)$}\\
        \item $\pmb{\hat{s'}} := \mathtt{NTT} (\pmb{s'}) \in \textcolor{black}{R_q^{(\ell)}}$ \\
        \item $\pmb{\hat{b'}} := \pmb{\hat{A}}^T\circ \pmb{\hat{s'}}$\\
        \item $\pmb{b'} := ( \mathtt{INTT}(\pmb{\hat{b'}}) + \pmb{e'}) \in \textcolor{black}{R_q^{(\ell)}}$\\
         \item $ \hat{c_m} := \pmb{\hat{b}}^T\circ \pmb{\hat{s'}}$
        \item $ c_m := \mathtt{INTT}(\hat{c_m}) + {e''} + \mathtt{Encode}(m) \in \textcolor{black}{R_q}$\\
        \item $\pmb{u} := \mathtt{Compress}(\pmb{b'},\  p)  \in \textcolor{black}{R_p^{(\ell)}}$\\
        \item $v := \mathtt{Compress}(c_m,\  t+2^B) \in R_{t+2^B} $\\
        \item \textbf{return} $c :=(\pmb{u},\  v)$
    \end{enumerate} 
    \end{small}
\end{subfigure}
\begin{subfigure}[t]{0.47\textwidth}
    \raggedright
    \vspace{-20mm}
    \begin{small}
    $\mathtt{PKE}{.}\mathtt{Dec}(\textcolor{black}{\mathrm{sk}}:=\textcolor{black}{\pmb{\hat{s}}},\ c:=(\pmb{u},\  v))$
    \begin{enumerate}[wide=0em, itemsep=0pt, parsep=0pt, font=\scriptsize\tt\color{gray}]
        \item $\pmb{u'} := \mathtt{Decompress}(\pmb{u},\  p)  \in \textcolor{black}{R_q^{(\ell)}}$\\
        \item $v' := \mathtt{Decompress}(v,\  t+2^B) \in R_{q} $\\
        \item $\pmb{\hat{u'}} := \mathtt{NTT} (\pmb{u'}) \in \textcolor{black}{R_q^{(\ell)}}$ \\
        \item ${\hat{u''}} := \pmb{\hat{u'}}^{\textcolor{black}{T}}\circ \textcolor{black}{\pmb{\hat{s}}} \in \textcolor{black}{R_q}$ \\
        \item $m'' := v' - \mathtt{INTT}({\hat{u''}}) \in \textcolor{black}{R_q}$ \\
        \item $m' := \mathtt{Decode}(m'') \in R_{2^B}$  \\
        \item \textbf{return} $m'$
    \end{enumerate} 
    \end{small}
\end{subfigure}
\end{varwidth}}
\vspace{-5pt}
\caption{MLWE based IND-CPA secure PKE using NTT}
\label{fig:kyberpke}
\vspace{-10pt}
\end{figure}

A generic MLWE-based public-key encryption (PKE) is shown in Fig.~\ref{fig:kyberpke}. It consists of three algorithms: (i) key-generation (\texttt{PKE.KeyGen}) generates public-key {$\mathrm{pk}$} and secret-key {$\mathrm{sk}$}, (ii) encryption (\texttt{PKE.Enc}) takes the public-key {$\mathrm{pk}$} and message $m$ as inputs and generates ciphertext $c$, and (iii) decryption (\texttt{PKE.Dec}) takes inputs as ciphertext $c$ and secret-key {$\mathrm{sk}$} and recovers the encrypted message. This PKE scheme is indistinguishable under chosen plaintext attacks (IND-CPA) based on the assumption of the hardness of the MLWE problem.  Here, $q$ is a prime modulus, and $p,\ t$ are power-of-2 moduli. These algorithms use NTT to perform polynomial multiplication efficiently~\cite{patrick_longa_ntt}. 
{The $\mathtt{Compress}: R_q \longrightarrow R_p$ function is defined as $\mathtt{Compress}(x')=\frac{px'+\lfloor q/2\rceil}{q}{\bmod p}$. $\mathtt{Decompress}: R_p \longrightarrow R_q$ is defined as $\mathtt{Decompress}(x) = \lfloor \frac{q}{p}\rceil x$.}
The $\mathtt{Encode}: {R_{2^B}}\longrightarrow {R_q}$ is defined as $\mathtt{Encode}(m) = \lfloor \frac{q}{2^B}\rceil m$ and the $\mathtt{Decode}: {R_q}\longrightarrow {R_{2^B}}$ is defined as $\mathtt{Decode}(m'')=\frac{2^Bm''+\lfloor q/2\rceil}{q}{\bmod 2^B}$. {Compress, Decompress, Encode, and Decode operations are applied {coefficient-wise} to each polynomial and vector of polynomials.}

\subsection{MLWE-based Key Encapsulation Mechanism}\label{lwe-kem}
\begin{figure}[!b]
\vspace{-10pt}
\centering
\scriptsize
\fbox{\begin{varwidth}{\textwidth}
\begin{subfigure}[t]{0.50\textwidth}
    \raggedright 
    \begin{small}
    $\mathtt{KEM}{.}\mathtt{KeyGen} ()$
    \begin{enumerate}[wide=0em, itemsep=0pt, parsep=0pt, font=\scriptsize\tt\color{gray}]
        \item $(\textcolor{black}{\mathrm{pk}} := (\textcolor{black}{\mathrm{seed}}_{\pmb{A}},\  \pmb{\hat{b}})$
         $\textcolor{black}{\mathrm{sk}} := (\pmb{\hat{s}}))  := \mathtt{PKE}{.}\mathtt{KeyGen} ()$ \\
        \item $\textcolor{black}{\mathrm{pkh}} := \mathcal{H}(\textcolor{black}{\mathrm{pk}}) \textcolor{black}{\in \{0,\ 1\}^{\text{len}_{K}}}$ \\
        \item $z \leftarrow  \mathcal{U}(\{0,\ 1\}^{\text{len}_{K}}$) \\
        \item \textbf{return} $(\textcolor{black}{\overline{\mathrm{pk}}} := \textcolor{black}{{\mathrm{pk}}} = (\textcolor{black}{\mathrm{seed}}_{\pmb{A}},\  \pmb{\hat{b}}),\ \textcolor{black}{\overline{\mathrm{sk}}} :=  (\pmb{\hat{s}},\ z,\  \textcolor{black}{\mathrm{pkh}},\ \textcolor{black}{\overline{\mathrm{pk}}}))$
    \end{enumerate} 
    \end{small}
\end{subfigure}
\begin{subfigure}[t]{0.50\textwidth}
    \raggedright
    \begin{small}
    $\mathtt{KEM}{.}\mathtt{Encaps} (\textcolor{black}{\overline{\mathrm{pk}}} := (\textcolor{black}{\mathrm{seed}}_{\pmb{A}},\  \textcolor{black}{\pmb{\hat{b}}}))$
    \begin{enumerate}[wide=0em, itemsep=0pt, parsep=0pt, font=\scriptsize\tt\color{gray}]
        \item \textcolor{black}{$\mathrm{msg}  \leftarrow  \mathcal{U}(\{0,\ 1\}^{\text{len}_{K}})$} \\
        \item \textcolor{black}{$m := \mathtt{Arrange\_msg}(\mathrm{msg}) \in R_{2^B}$} \\
        \item $(K,\  r) := \mathcal{G}(\mathcal{H}(\textcolor{black}{\overline{\mathrm{pk}}}),\  m) \textcolor{black}{\in \{0,\ 1\}^{\text{len}_{K}} \times \{0,\ 1\}^{\text{len}_{K}}}$ \\
        \item $c := \mathtt{PKE}{.}\mathtt{Enc} (\textcolor{black}{\overline{\mathrm{pk}}},\  m;\  r )$ \\
        \item \textbf{return} $(c,\  K)$
    \end{enumerate} 
    \vspace{2mm}
    \end{small}
\end{subfigure}
\begin{subfigure}[t]{\textwidth}
    \raggedright
    \begin{small}
    $\mathtt{KEM}{.}\mathtt{Decaps} (\textcolor{black}{\overline{\mathrm{sk}}} := (\pmb{\hat{s}},\  z,\  \textcolor{black}{\mathrm{pkh}},\ ,\ \textcolor{black}{\overline{\mathrm{pk}}}),\ c)$
    \begin{enumerate}[wide=0em, itemsep=0pt, parsep=0pt, font=\scriptsize\tt\color{gray}]
        \item $m'  :=  \mathtt{PKE}{.}\mathtt{Dec} (\textcolor{black}{\pmb{\hat{s}}},\  c )$ \\
        \item \textcolor{black}{$\mathrm{msg'} := \mathtt{Original\_msg}(\mathrm{m'}) \in \{0,\ 1\}^{\text{len}_{K}}$} \\
        \item $(K',\  r') := \mathcal{G}(\textcolor{black}{\mathrm{pkh}},\  m') \textcolor{black}{\in \{0,\ 1\}^{\text{len}_{K}} \times \{0,\ 1\}^{\text{len}_{K}}}$ \\
        \item \textcolor{black}{$m'' := \mathtt{Arrange\_msg}(\mathrm{msg'}) \in R_{2^B}$} \\
        \item $c_* := \mathtt{PKE}{.}\mathtt{Enc} (\textcolor{black}{\textcolor{black}{\overline{\mathrm{pk}}}},\  m'';\  r')$ \\
        \item $K'' := \mathcal{H}(c,\ z)$\\
        \item \textbf{if} $c=c_*$ \textbf{then} \textbf{return} $ K := K'$ \\
        \item \textbf{else} \textbf{return} $ K := K''$\\
    \end{enumerate} 
    \end{small}
\end{subfigure}
\end{varwidth}}
\vspace{-5pt}
\caption{MLWE based IND-CCA secure KEM using NTT}
\label{fig:kyberkem}
\vspace{-10pt}
\end{figure}
The PKE scheme described in Sec.~\ref{lwe-pke} is  IND-CPA. Indistinguishability under adaptive chosen ciphertext attack (IND-CCA) is a stronger security notion than IND-CPA and is desired to construct a KEM. The IND-CPA PKE in Fig.~\ref{fig:kyberpke} is converted to IND-CCA KEM by applying a variant of Fujisaki–Okamoto (FO) transformation~\cite{DBLP:conf/tcc/HofheinzHK17}. As the PKE scheme is based on the MLWE problem, the PKE scheme is not perfectly correct (when the decryption of the encrypted message does not return the original message). If the underlying PKE is $(1-\delta)${-}correct{,} then the KEM based on the PKE is also $(1-\delta)${-}correct~\cite{DBLP:conf/tcc/HofheinzHK17}. 
{Jiang et al.~\cite{Jiang2017} proposed a {IND-CCA} KEM construction from a {IND-CPA} $(1-\delta)${-}correct PKE in the quantum {random} oracle model, and a slightly modified version of it is used in {FrodoKEM}~\cite{DBLP:conf/ccs/BosCDMNNRS16}. The KEM shown in Fig.~\ref{fig:kyberkem} closely follows the {FrodoKEM} construction.}
The IND-CCA MLWE-based KEM consists of three algorithms{:} (i) key-generation (\texttt{KEM.KeyGen}), (ii) encapsulation (\texttt{KEM.Encaps}), and (iii) decapsulation (\texttt{KEM.Decaps}). These algorithms use two hash functions, namely $\mathcal{G}:\{0,\ 1\}^*\longrightarrow\{0,\ 1\}^{2*\text{len}_K}$ and $\mathcal{H}:\{0,\ 1\}^*\longrightarrow\{0,\ 1\}^{\text{len}_K}$. 
 {We also used two other functions. $\mathtt{Arrange\_msg}: \{0,\ 1\}^{\text{len}_{K}} \longrightarrow R_{2^B}$ is defined by $\mathtt{Arrange\_msg}(\mathrm{msg})=m$, where each of the coefficients of $m$ consists of $B$ bits of $\mathrm{msg}$. If ${n}>{\text{len}_{K}}$ and $\frac{n}{\text{len}_{K}}=\texttt{repeat}>1$, then \texttt{repeat} coefficients of $m$ consist of the same bit of $\mathrm{msg}$. $\mathtt{Original\_msg}:  R_{2^B} \longrightarrow \{0,\ 1\}^{\text{len}_{K}}$ is the inverse function of $\mathtt{Arrange\_msg}$. Therefore, $\mathtt{Original\_msg}(\mathtt{Arrange\_msg}(\mathrm{msg}))= \mathrm{msg}$. These two functions work as a repetition code with a majority vote when $\frac{n}{\text{len}_{K}}>1$ and random coin flip in case of ties.} We discuss {these} in more detail in Sec.~\ref{sec:module_space_exploration}.

\vspace{-15pt}
\subsection{Related works}
Most {of the current} lightweight PKC schemes are based on ECC~\cite {5537064,inproceedings_ecc,ECC_compact_1, ECC_compact_2} which are not secure against quantum adversaries. Lattice-based constructions are promising candidates for designing lightweight PQC schemes. The NIST standard lattice-based KEM (e.g. Kyber~\cite{Kyber-Kem}) or the finalists of NIST standardization (e.g. Saber~\cite{Saber_kem}) are mainly designed keeping security and performance in mind. 
{Afterward, LW implementations of these schemes {have been} proposed.}
For reference, {Huang et al.}~\cite{DBLP:journals/ieiceee/HuangHLW20}, Xing {and Li}~\cite{compact_kyber}, Ni et al.~\cite{Ayesha-k2-reduction} proposed optimized implementations of Kyber in various hardware platforms. Roy {and Basso}~\cite{DBLP:journals/tches/RoyB20} presented an implementation of Saber on FPGA hardware, and Ghosh et al.~\cite{DBLP:conf/cicc/GhoshMKDGVS22,DBLP:journals/jssc/GhoshMKDGVS23} proposed an area and energy-optimized implementation of Saber in ASIC. There are several hardware or hardware/software co-designs of Kyber available~\cite{DBLP:journals/iacr/BasuSNK19,DBLP:journals/iacr/DangFAMNG20,DBLP:journals/iacr/BanerjeeUC19}.    

{An RLWE-based encryption scheme with binary secrets and errors, called Ring-BinLWE, was proposed in~\cite{DBLP:conf/ccs/BuchmannGGOP16} suitable for lightweight PKC applications.}
Subsequently, more efficient variants of this scheme have been published in the following works~\cite{9737700,9586151,8660431,9415642}. However, these schemes are only IND-CPA secure and hence vulnerable to the chosen ciphertext attacks (CCA). Later on, Ebrahimi et al.~\cite{9027941} proposed a CCA secure version of the IND-CPA Ring-BinLWE scheme, but the quantum bit security provided by this scheme is $<75$. This is relatively lower in comparison to Saber and Kyber, which provide at least 100-bit {core-SVP} PQ security even in their lowest security versions.
There is always a trade-off between efficient implementation in the resource-constrained platform and security, and not much work has {been devoted towards} designing lightweight PQC without compromising security.     

During the NIST PQC standardization procedure, a suite consisting of three learning with rounding (a variant of LWE problem) based PQC KEMs, Scabbard, was proposed {in}~\cite{scabbard}. 
This work explored new design choices, such as a small polynomial size {with} $n=64$ for one of the schemes (Espada) to reduce the memory footprint of the implementation {on} the resource constraint Cortex-M4 device. Before that, the smallest polynomial size $n$ used in (R/M)LWE-based KEM was $256$. In addition to this, several new designs of LWE-based KEMs, such as Smaug~\cite{smaug_kem}, TiGER~\cite{TiGER}, etc., have been submitted in the ongoing Korean PQC competition~\cite{kpqc}.
Although the aforementioned works improved the state-of-the-art of LBC with different design choices and implementations, none of them explored all the possible design choices of LWE-based KEMs from the perspective of lightweight hardware implementations.   

\vspace{-10pt}
\section{Rudraksh: Design Space Exploration}\label{sec:our_design}
Designing a cryptographic scheme is fundamentally solving {a multidimensional} optimization problem where the primary objective functions are attaining a particular level of security, reducing latency and bandwidth (the size of the public key, ciphertext, and secret key). However, we impose new constraints on our lightweight design, such as low memory, low energy, and low area requirements {to execute} the scheme with reasonable latency. Our LBC design is influenced primarily by $3$ parameters, the structure of the module \textit{i.e} the rank of the matrix {$\ell$} and the size of the constituent polynomials $n$, the prime modulus $q$, and the standard deviation $\sigma$ of the secret or error distribution. In this section, we discuss our design decisions and the rationale behind them in choosing these variables to achieve our design objective of a lightweight KEM.

\subsection{Module space exploration}\label{sec:module_space_exploration}
\begin{figure}[!ht]
\vspace{-10pt}
  \centering
  \includegraphics[scale=.48]{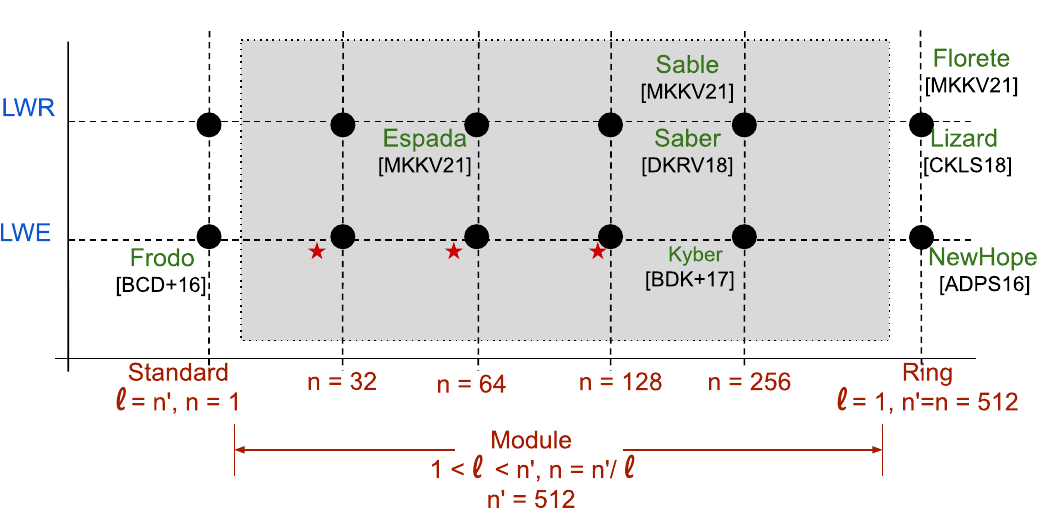}
  \vspace{-10pt}
  \caption{{Design space of lattice-based KEMs depending on the different variations of LWE problems and underlying ring sizes. Our explored module spaces are marked with red stars.}}
  \label{fig:design_n}
  \vspace{-10pt}
\end{figure}

{Keeping the $q$ and $\sigma$ fixed, the security of the standard, module, or ideal lattice-based cryptographic scheme is dependent on the dimension $n'$ of the underlying lattice only.}
Module lattices present a convenient and generic representation of different lattices; therefore, in this section, we will use the modules to describe different types of lattices. {Let us} consider a {square} module lattice\footnote{{$\mathbf{A}$ in (M/R)LWE-based KEMs (or (M/R)LWR-based KEMs) is a square-matrix (i.e., number of rows = number of columns) to ensure the same security of key-generation (which uses $\mathbf{A}$) and encapsulation (which uses $\mathbf{A^T}$) as described in~\cite{DBLP:conf/ctrsa/LindnerP11}.
}} $\pmb{A}\in {R_q^{({\ell \times \ell})}}$. The rank of the underlying {lattice} is $n'={\ell}\times n$. 
Upon fixing $n=1$ and ${\ell}=n'$, we get a standard lattice. On the other hand, if we fix ${\ell}=1$ and $n'=n$ \textit{i.e.} our lattice consists of a single polynomial, and we get an ideal lattice ({RLWE}). 
Indeed, for a long time before the proposal of module lattices~\cite{module_sis}, these two extremities were the only two choices available to design lattice-based cryptosystems, as shown in Fig.~\ref{fig:design_n}. Although Kyber~\cite{Kyber-Kem}, Saber~\cite{Saber_kem}, and Dilithium~\cite{dilithium} are prominent examples of cryptographic schemes based on module lattices, a vast spectrum of lattice configurations with different values of {$\ell$} or $n$ {have} been left unexplored. This is shown in Fig.~\ref{fig:design_n} as a grey-shaded region.
We explore this region to find optimal choices for $n$ and {$\ell$} to design a lightweight KEM. It should be noted that this is not trivial. Intuitively, one might think that choosing a small $n$ would immediately lead to a lightweight design as it reduces the size of the multiplier. However, to maintain the $n'$ for the security, decreasing $n$ increases {$\ell$}. This implies more multiplications, more random numbers, larger moduli, etc.
Similarly, just decreasing {$\ell$} is also not useful for LW designs. We have to strike a delicate balance between {$\ell$} and $n$ and other metrics that influence the scheme's suitability for small resource-constrained devices. We discuss these different metrics and how they are affected by different values of $n$ and {$\ell$} below.

\noindent
\textbf{Memory consumption: }The matrix-vector multiplication is performed in \texttt{PKE.KeyGen} (therefore in \texttt{KEM.KeyGen}) and \texttt{PKE.Enc} (therefore in \texttt{KEM.Encaps} and \texttt{KEM.Decaps}) algorithm (shown in Fig.~\ref{fig:kyberpke}). The storage requirement for the public-matrix $\pmb{A}$ is one of the most memory-expensive operations for the LWE {or LWR} schemes that use module lattice structure. It requires storing ${\ell \times \ell}$ polynomials of degree $n-1$. Currently, the \textit{de-facto} standard of lattice-based implementation is to generate this matrix using the \textit{just-in-time}~\cite{saber_on_arm} strategy. This method generates the matrix $\pmb{A}$ one polynomial at a time by utilizing the sponge-based {periodic} `..squeeze-absorb-squeeze..' operation of the extended output function (XOF) such as Keccak~\cite{keccak_ec}. Therefore, the memory requirement to perform the matrix-vector multiplication is proportional to the size of one single polynomial. As we move towards the left of Fig.~\ref{fig:design_n}, polynomial size $n$ decreases. So, although {$\ell$} has to be increased to maintain the security level, the memory requirement in this configuration is smaller. We store a single polynomial for all the polynomial multiplications in hardware platforms. Of course, one can take extreme measures such as generating a single coefficient at a time and performing a single integer multiplication to reduce memory. However, it would drastically deteriorate performance. 

{In the MLWR-based schemes, the error vectors are generated implicitly. This implies that we do not need to invoke the XOF (or the CBD) module to generate the error vectors in the MLWR-based schemes. Nevertheless, we need to invoke these modules to generate secret vectors in both MLWR and MLWE-based schemes. Therefore, we need these modules on the hardware in either of these schemes. Hence, the implicit generation of error vectors in MLWR-based schemes offers no advantage from area-optimization point-of-view.
Also, for the rounding operation, in MLWR-based schemes such as Saber~\cite{Saber_kem}, the modulus and the rounding modulus are chosen as power-of-two numbers. Due to this, MLWR-based KEMs use Toom-Cook-based polynomial multiplication, which generally consumes a relatively larger area/power/latency than NTT-based multiplication used in MLWE-based schemes~\cite{scabbard-tecs}. In MLWE-based KEMs that support NTT natively, $\pmb{\hat{A}}\in R_q^{(\ell \times \ell)}$ can be sampled directly from the NTT domain instead of first sampling $\pmb{A}$ and then performing NTT($\pmb{A}$) to generate $\pmb{\hat{A}}$. However, one can also perform the evaluation stage of Toom-Cook multiplication (evaluation-schoolbook multiplication-interpolation) and store it~\cite{Saber_Time-memory,polymul-z2mx-m4}. However, it is not in-place like NTT and requires more memory than the original $\pmb{A}$. Although following the work of Chung et al.~\cite{tches:ChungHKSSY21}, an NTT can be used for MLWR-based schemes by choosing a large NTT-friendly prime that envelopes the modulus and growth of error during multiplications. 
This strategy helps MLWR-based schemes, such as Saber, achieve performance comparable to Kyber on the ARM Cortex-M4 platform. However, the area requirement (or the stack memory of Cortex-M4 implementations~\cite{Abdulrahman_Chen_Chen_Hwang_Kannwischer_Yang_2021}) increases significantly for MLWR-based KEMs due to their larger modulus.
Recently, three MLWR-based KEMs have been proposed in~\cite{scabbard} and their hardware implementation in~\cite{scabbard-tecs}. This work shows that Toom-Cook-based polynomial multiplication can be made resource-constrained at the cost of more cycles. 
Therefore, we explore possible MLWE schemes with smaller polynomial sizes, which can be implemented with low resources without much performance degradation.}
We primarily target schemes where $n$ is power-of-2 and less than $256$, such that $128$, $64$, and $32$.  

\noindent
\textbf{Multiplier size: }In the case of a standard lattice-based scheme with matrix rank $n'$, one of the most computation-heavy operations is multiplications between $n'\times n'$ matrix and $n'$ length vector. For the ring lattice-based scheme, we have to perform polynomial multiplications between two $n'-1$ degree polynomials to achieve a similar level of security. This can be done using quasi-linear NTT multiplication. Hence, for a particular security level, the ring lattice-based schemes are more efficient than the standard lattice-based schemes in terms of computational cost. However, the resource consumption in that case is relatively huge as two $n'-1$ degree polynomials must be stored to perform the polynomial multiplication. Therefore, for a specific security level, due to the \textit{just-in-time} strategy, the module lattice-based schemes are more beneficial in reducing resource consumption than the ring lattice-based schemes. Although we have to perform multiple polynomial multiplications due to the use of module structure, module lattice-based schemes perform better than the standard lattice-based schemes. The choice of the hard problem \textit{i.e.} MLWE or MLWR, and polynomial size determines the size of the multiplier in hardware. Nevertheless, the resource consumption is proportional to the size of a single polynomial for module lattice-based schemes. Therefore, choosing the hard problem and polynomial size is one of the leading factors when designing an efficient scheme for resource-constrained devices.  
{Generally, the size of the polynomial $n$ is chosen in multiple with the size of the secret message ($\mathrm{msg}$) bit-length $\text{len}_K$~\cite{newhope,Kyber-Kem,lizard}. If $n>\text{len}_K$, $B=1$, and $\frac{n{\cdot}B}{\text{len}_K}={\texttt{repeat}}$  for an integer ${\texttt{repeat}>1}$, we use ${\texttt{repeat}}$ coefficients of ciphertext polynomial $v$ (generated during {the} encryption algorithm shown in Fig.~\ref{fig:kyberpke}) to hide a single message bit of $\mathrm{msg}$ by replicating a single message-bit ${\texttt{repeat}}$ times using $\mathtt{Arrange\_msg}$ function. {Please note that we use $\texttt{repeat}>1$ of the message bits only when $n>\text{len}_K$, otherwise $\texttt{repeat}=1$.} 
{We now discuss the process of calculating failure probability when the \texttt{repeat} is greater than 1.}
{For example, if \texttt{repeat}=3, the failure happens when at least two of the three message bits are decoded to the wrong value. Then the new failure probability is calculated as $(3\cdot(\mathtt{fail\_prob})^2\cdot(1-\mathtt{fail\_prob}))+(\mathtt{fail\_prob})^3$, where $\mathtt{fail\_prob}$ is the failure probability when \texttt{repeat}=1 (no repetition).} 
{If \texttt{repeat}=4, the failure occurs when at least three of the four message bits are decoded to the wrong value. Failure also occurs when two of the four message bits are wrong with half probability (as in this case, one of the two choices would be chosen randomly, which can be correct with only half probability).} 
{Then the new failure probability is calculated as $(3\cdot(\mathtt{fail\_prob})^2\cdot(1-\mathtt{fail\_prob})^2)+(4\cdot(\mathtt{fail\_prob})^3\cdot(1-\mathtt{fail\_prob}))+(\mathtt{fail\_prob})^4$.}
{This calculation changes depending on \texttt{repeat}. However, we would like to mention that a designer can choose various approaches to encode message bits into the ciphertext polynomial when $n>\text{len}_K$.} Unlike the error correcting code used in LAC~\cite{DBLP:journals/iacr/LuLZJXHL18} where timing attack has been shown in~\cite{pkc-2019-29313}, the $\mathtt{Arrange\_msg}$ (and also $\mathtt{Original\_msg}$) function can be implemented in a constant-time manner. Now,} {if} the polynomial-size $n$ is smaller than the size of the message bit-length $\text{len}_K$ \textit{i.e.} $n=(1/B){\cdot}\text{len}_K$ for an integer $B>1$. {In that case,} we have to encode $B$ message bits into a single coefficient of ciphertext polynomial $v$~\cite{DBLP:conf/ccs/BosCDMNNRS16} as displayed in Fig.~\ref{fig:kyberpke}. 
This would increase the requirement of the reconciliation bits ($\log_{2}{t}$) and eventually the modulus of a coefficient of $v$ ($=\log_{2}{t}+B$). This will require a larger modulus $q$, which reduces the security. We will discuss this phenomenon in more detail in Sec.~\ref{sec:moduli_choice}. 

\noindent
\textbf{Flexible design: } Module lattices present an opportunity to design cryptographic schemes that benefit the IoT architecture as discussed in Sec.~\ref{sec:intro}.
The structure of module lattices can be utilized to instantiate such a flexible scheme. {Let us assume} a cryptographic scheme uses a module lattice of size ${\ell \times \ell}$. An extremely low latency and high resource-consuming implementation can be realized by implementing ${\ell^2}$ multipliers in parallel, and an extremely lightweight and high latency version can be realized by implementing a single multiplier repeatedly using for ${\ell^2}$ times. Here, of course, the XOF has to be implemented accordingly to match the latency of the multiplier.
As discussed before, polynomial arithmetic, specifically polynomial multiplication, is one of the major bottlenecks in LBC's performance and resource consumption. The polynomial size in both Kyber and Dilithium is $256$. Even in the most lightweight instantiation, an IoT peripheral device has to use a $256\times 256$ polynomial multiplier {(more specifically, NTT multiplication, which includes size-$256$ NTT, size-$256$ INTT and size-$256$ point-wise multiplication)}. This is still very expensive for an IoT peripheral device. A smaller polynomial size is more suitable with some sacrifice in efficiency. Therefore, a balance has to be struck between these two metrics for a suitable lattice-based PQ scheme for IoT. 

In conclusion, ring lattice-based schemes are especially advantageous for achieving better performance{,} but {they require} more hardware area. Meanwhile, standard lattice-based schemes theoretically can be implemented with lesser memory and area at the expense of substantial computation costs. Module lattice-based schemes with polynomial-size $n$ and underlying lattice's matrix rank $n'={\ell}\times n$ provide a trade-off between them. If we keep $n'$ constant, {and increase} {$\ell$} then $n$ {decreases proportionately}, which reduces the memory requirements for a single polynomial. However, this increases the generation cost of the matrix $\pmb{A}$, as the matrix consists of ${\ell}\times {\ell}\times n\times \lceil \log_2q\rceil$ pseudo-random bits. These pseudo-random numbers are generated by using {an} XOF, which is another computationally expensive operation as described later in Sec.~\ref{sec:ASCON_xof}. If memory is not a concern, then increments of {$\ell$} can be used to increase parallelism in hardware implementations. There are some module LWR-based designs that have been proposed in recent years~\cite{Saber_kem,scabbard}. However, the module-space for designing different MLWE-based schemes remained mostly unexplored. Therefore, we choose to explore the module space for the LWE problem denoted by red colored stars in Fig.~\ref{fig:design_n} to construct a lightweight MLWE-based KEM with optimal parameters. 

\vspace{-10pt}
\subsection{Choice of moduli}\label{sec:moduli_choice}
It is clear from the discussion in the previous section that we want to explore the module lattice space to design LW lattice-based KEM.
LWE-based schemes use reconciliation mechanisms by sending some extra bits~\cite{Kyber-Kem} {($t$ in Fig.~\ref{fig:kyberpke})}. These bits help to recover the encrypted message during the decryption algorithm by reducing the noise introduced during the encryption procedure called decryption noise (as LWE-based encryption schemes are not perfect). Increased decryption noise induces an increment in the failure probability, which can cause a decryption failure attack~\cite{pkc-2019-29313}. 
As discussed earlier, a smaller polynomial size $n$ reduces the memory consumed by a single polynomial and also the area required to implement single polynomial multiplication in hardware. But, if we reduce the size of the polynomial $n$, then we have to encode multiple message bits in a coefficient of $v$. This will increase the failure probability. This can be compensated by more reconciliation bits {i.e., larger $t$}, which in turn increases $q$ {and the ciphertext size}.

The security of a lattice-based cryptosystem increases with the increase in the error-to-modulus ratio, \textit{i.e.} keeping the error distribution fixed, the security will reduce with the increase in the value of $q$, and vice versa.
Therefore, a smaller value of $q$ helps to increase efficiency, reduce computational and storage resources, and also reduce the bandwidth (size of the public-key, secret-key, and ciphertext), but increases the failure probability.
Hence, we concentrated on finding an optimal value of the modulus $q$ for which the decryption failure would be minimal. We also proposed implementations with minimal hardware resources.
The type of the modulus \textit{i.e} prime vs. power-of-$2$ modulus also has a significant impact on the performance and resource utilization of the scheme. We have deferred this discussion till Sec.~\ref{sec:design_NTT}. 

\subsection{Choice of polynomial multiplication}\label{sec:design_NTT}
For LBC schemes, polynomial multiplication is one of the major bottlenecks with respect to efficiency and resource consumption. In literature, there exist mainly two types of polynomial multiplication algorithms for implementing LBC schemes: i) Toom-Cook multiplication~\cite{toom,cook}, and (ii) NTT multiplication~\cite{Pollard1971TheFF,patrick_longa_ntt}. Toom-Cook multiplication is relatively simpler and can be used for any modulus. However, the time complexity of the Toom-Cook polynomial multiplication is asymptotically slower $O(n^{1+\epsilon})$, where $0<\epsilon<1$. On the other hand, NTT multiplication is the most used polynomial multiplication for implementing LBC schemes due to its faster quasi-linear time complexity ($O(n\log_2 n)$). However, for the NTT multiplication, the modulus $q$ needs to be \emph{NTT friendly} \textit{i.e.} a prime number with the primitive $2n$-th {root-of-unity} in the prime field $\mathbb{Z}_q$. Please note that, for small values of $n$, the efficiency of Toom-Cook polynomial multiplication with a power-of-$2$ modulus (as modular reduction is free in a power-of-$2$ ring) and NTT-based polynomial multiplication on an appropriate prime modulus is comparable when performed the full multiplication. 
However, for LBC schemes, we can sample the public matrix $\mathtt{NTT}(\pmb{A})= \pmb{\hat{A}}$ from ${R_q^{(l\times l)}}$ using $seed_{\pmb{A}}$ instead of sampling $\pmb{A}$ and {performing} NTT on $\pmb{A}$. It can save cycles while computing $\pmb{A}\cdot \pmb{s}$ as $\pmb{A}$ is random implies ${\mathtt{NTT}}(\pmb{A})$ is random and vice-versa. We also can save execution time on the NTT-based polynomial multiplication by omitting the INTT operation and keeping the multiplication result in the NTT domain (e.g., Line (5) in the PKE.KeyGen() in Fig.~\ref{fig:kyberpke}). This makes the LBC scheme with NTT multiplication more efficient than Toom-Cook multiplication. Therefore, we choose to use NTT-based polynomial multiplication over a prime {modulus} $q$.

NTT multiplication between two polynomials $a$ and $b$ from {$R_q$} is performed by $a\cdot b = \mathtt{INTT}(\mathtt{NTT}(a)\circ \mathtt{NTT}(b))$.
Given $\zeta$ is the $2n$-th primitive root of unity and $\omega = \zeta^2$, the $\mathtt{NTT}(x) = \hat{x} = {(\hat{x}_0,\ \hat{x}_1,\ \ldots, \hat{x}_{n-1})}$ and $\mathtt{INTT}(\hat{x}) = {x} = ({x_0},\ {x_1},\ \ldots, {x_{n-1}})$ are denoted by the following Eq.~\ref{eq:ntt}~\&~\ref{eq:intt}.
\begin{align}
    \hat{x_i} &= \sum_{j=1}^{n-1} x_j \zeta^{(2i+1)j} = \sum_{j=1}^{n-1} (x_j\zeta^{j}) \omega^{ij} \bmod{q},\ 0\leq i \leq n-1.\label{eq:ntt}\\
    {x_j} &= 1/n \sum_{i=1}^{n-1} \hat{x_i} \zeta^{-(2i+1)j} = \zeta^{j}/n \sum_{i=1}^{n-1} \hat{x_i} \omega^{-ij} \bmod{q},\ 0\leq j \leq n-1. \label{eq:intt} 
\end{align}
In this procedure of multiplication, we have to store the pre-computed values of $\zeta^{j} \bmod{q}$ (for $1\leq j\leq n-1$) along with the coefficients of two participated polynomials for improving the performance. Therefore, the total memory requirement to perform multiplication depends on polynomial size $n$.

In the literature, there is another type of NTT called incomplete NTT multiplication, where the NTT multiplication between two $n$ size polynomials is replaced by two separate NTT multiplications between two $n/2$ size polynomials. Karatsuba multiplication is performed on the last $1$-degree polynomials. Therefore, incomplete NTT multiplication requires more modular multiplications than complete NTT multiplication. Yet, the incomplete NTT multiplications outperform complete NTT multiplications~\cite{Alkim_Alper_Bilgin_Cenk_Gérard_2020} on software by omitting several reduction steps after modular multiplication. Incomplete NTT multiplication is also used in Kyber. We employ a single butterfly module shown in Fig.~\ref{fig:butterfly} that includes a reduction step for performing NTT, INTT, and PWM in our KEM. Therefore, incomplete NTT multiplication increases the latency in our KEM implementation. Consequently, we choose the complete NTT multiplication over the incomplete one for polynomial multiplication. More details regarding NTT multiplication are provided in Sec~\ref{sec:butterfly_hw}.

\subsection{Secret and error distribution}
In LWE-based schemes, coefficients of secret and error are usually sampled from a narrow distribution. There are several (M/R)LWE-based KEMs that have utilized discrete Gaussian distribution as secret and error distribution~\cite{DBLP:conf/ccs/BosCDMNNRS16}. Unfortunately, it is hard to implement a Gaussian sampler efficiently and securely against timing attacks. For KEMs, {Alkim et al.}~\cite{newhope} showed that this Gaussian distribution can be replaced with a centered binomial distribution (CBD) whose standard deviation is the same as the Gaussian distribution. The sampling from a CBD is much simpler and easier to protect against side-channel attacks.
Several other distributions have been explored in the design of LWE-based schemes to gain efficiency, such as binary distribution~\cite{DBLP:conf/ccs/BuchmannGGOP16}, fixed weight distribution~\cite{round5, smaug_kem, TiGER}, etc. 
{However, as our goal was to design a Kyber\textit{-esque} scheme, we had limited our search space to CBDs ($\beta_\mu$) with different $\mu$. The above-mentioned other distributions can also be potentially used as secret and error distributions for lightweight cryptography, but we have not investigated them in this work.}

{The parameter $\mu$ impacts the security parameters of a CCA secure scheme, failure probability, and bit security. The standard deviation of $\beta_\mu$ is $\sqrt{\frac{\mu}{2}}$. Suppose the modulus $q$ and the rank of the lattice $n'$ are fixed. In that case, {both the bit security and the failure probability of the scheme increase} as the standard deviation (for CBD, $\mu$) increases.} 
The sampling from a CBD $\beta_\mu$ is accomplished by performing $\mathtt{HW}(a)-\mathtt{HW}(b)$, where $a,\ b $ are $\mu$ bit pseudo-random numbers. The parameter $\mu$ of CBD is crucial in deciding the scheme's efficiency. 
The CBD sampler uses pseudo-random numbers, and the bigger the CBD parameter $\mu$, the more pseudo-random numbers generation will be required. Pseudo-random numbers are generated using some extendable-output function (XOF). The XOF is one of the costliest operations in terms of computation and resources (Details about XOF have been provided in the next subsection). Finding a smaller $\mu$ is necessary for better performance, less resource utilization, and lower failure probability. However, a larger $\mu$ increases the bit security. As we are aiming for a CCA-secure KEM scheme with $100$-bit {Core-SVP} PQ security using FO transform~\cite{DBLP:conf/tcc/HofheinzHK17}, the failure probability must be $\leq 2^{-100}$.

Hence, in our design, we have analyzed these aspects with a wide range of values of $\mu$ to find an optimal choice to strike a balance between the security and efficiency of our scheme.

\subsection{ASCON based hash and XOF functions}\label{sec:ASCON_xof}
The implementations of LBC exhibit a unique and interesting phenomenon. Lattice-based cryptographic schemes use a lot of pseudorandom numbers to generate the matrix $\pmb{\hat{A}}$ and the secret $\pmb{s}$ and error $\pmb{e}$ vectors. From a designer's perspective, generating pseudorandom numbers is considered an auxiliary function that does not impose major overhead on executing the whole scheme. More focus is given to optimizing the core functions of the cryptographic scheme as they consume the majority of the time and resources in {various implementations.} This is true for classical PKC (and symmetric-key ciphers also) schemes such as RSA~\cite{RSA} and ECC~\cite{ECC_miller_Crypto86}, where most of the time and resources are spent on making the multi-precision multiplications and scalar point multiplication, respectively. However, for LBC, the standard approach of generating pseudorandom numbers is using an XOF such as Keccak~\cite{keccak_ec}. This process takes close to or, in some cases, more than $50\%$ of total time and/or area~\cite{compact_kyber}. As numerous works have been done to optimize the core operation of LBC, which is polynomial multiplication in software and hardware platforms~\cite{DBLP:journals/tches/RoyB20,DBLP:conf/dac/MeraTKRV20}, the process of random number generation has become the bottleneck.

To alleviate this problem, designers have proposed {alternative} versions of their schemes, such as Kyber-90s~\cite{kyber_specification} or Saber-90s~\cite{saber_specification_3rd_round} where they proposed to generate the random numbers using a block cipher (such as AES~\cite{AES_compact_1}) in counter mode. While this could be a good solution for software and hardware platforms with dedicated support for these block ciphers, such as the AES-NI instruction set, for standalone hardware with minimal software support, this is not a good solution. As shown in Fig.~\ref{fig:kyberkem} (for $\mathcal{G},\ \mathcal{H}$), we need to use {secure hash functions} for a CCA-secure KEM using {the} FO transformation. Therefore, we cannot completely remove the Keccak module from the hardware platform. Moreover, we must include another module implementing the block cipher algorithm. 
{Another strategy for reducing the overhead of Keccak could be using round-reduced Keccak~\cite{round-rdeuced-keccak} to generate $\pmb{\hat{A}}$, $\pmb{s}$, and $\pmb{e}$ ($\pmb{s'}$, $\pmb{e'}$, $e''$), where only uniform randomness is needed~\cite{google-forum-round-rdeuced-keccak}. However, there is not enough research to engender enough confidence to use the round-reduced Keccak as a reliable pseudo-random number generator. Therefore, more research is required to determine the security versus efficiency trade-off. Moreover, we wanted to use a \textit{standard} hash/XOF function, which went through several security analyses in our scheme.} 

Therefore, the best possible solution in this scenario is to replace the \textit{bulky} Keccak module with some lightweight alternative. NIST concluded its lightweight cryptography competition~\cite{NIST_LWC} in February 2023 and selected the ASCON~\cite{ASCON} family of lightweight ciphers. {Like Keccak}, ASCON is also based on the sponge construction~\cite{sponge_construction} and can be used as a Hash function and XOF. Moreover, ASCON is specifically designed for lightweight implementation on resource-constrained devices. This makes ASCON an ideal choice for replacing the Keccak function in our design. However, this is not very straightforward. The biggest hurdle is the difference in the state size of these two ciphers, which are $320$ and $1600$ {bits} for ASCON and Keccak, respectively. Therefore, each ASCON-squeeze outputs a fraction of pseudorandom bits compared to a Keccak-squeeze. Hence, to utilize ASCON's full potential, we have carefully designed our architecture exploiting the lightweight sub-layer and linear layer of ASCON, as well as meticulous scheduling and memory organization in the FPGA implementation {such that} the smaller throughput does not become the operational bottleneck (explained in Sec.~\ref{sec:ascon_hw}). Another hurdle in replacing Keccak with ASCON is that the current version of ASCON provides maximum 128-bit security; therefore, it is {unsuitable} to replace SHAKE-256 or SHA3-512, which has been used in Kyber {for higher security versions}. However, it is fine for our lightweight design. {Due to these issues, replacing Keccak with ASCON and achieving efficiency is not straightforward.}

\subsection{Parameters of our scheme} \label{sec:param}
\begin{table}[!t]
\vspace{-10pt}
\centering
\caption{Parameter set of all the explored designs of KEMs together with Kyber and NewHope for NIST-level-1 security.}
\label{tab:design_exploration_low}
\resizebox{\textwidth}{!}{%
\begin{tabular}{c|cc|cc|cc|cc|c|c|c}
\hline
{\color[HTML]{3531FF} }                                                                                  & \multicolumn{2}{c|}{{\color[HTML]{3531FF} \begin{tabular}[c]{@{}c@{}}Module\\ Parameter\end{tabular}}} & \multicolumn{2}{c|}{{\color[HTML]{3531FF} \begin{tabular}[c]{@{}c@{}}Primary \\ modulus\end{tabular}}} & \multicolumn{2}{c|}{{\color[HTML]{3531FF} \begin{tabular}[c]{@{}c@{}}Compression \\ modulus\end{tabular}}} & \multicolumn{2}{c|}{{\color[HTML]{3531FF} \begin{tabular}[c]{@{}c@{}}CBD \\ parameter\end{tabular}}} & {\color[HTML]{3531FF} Encoding} & {\color[HTML]{3531FF} {Bit-security}}             & {\color[HTML]{3531FF} }                                                                                \\ \cline{2-11}
\multirow{-3}{*}{{\color[HTML]{3531FF} \begin{tabular}[c]{@{}c@{}} {Scheme}\\{name}\\ \end{tabular}}} & {\color[HTML]{3531FF} {$\ell$}}                         & {\color[HTML]{3531FF} $n$}                        & {\color[HTML]{3531FF} $q$}                & {\color[HTML]{3531FF} $\lceil\log_2q\rceil$}               & {\color[HTML]{3531FF} $\lceil\log_2p\rceil$}         & {\color[HTML]{3531FF} $\lceil\log_2t\rceil$}        & {\color[HTML]{3531FF} $\eta_1$}                   & {\color[HTML]{3531FF} $\eta_2$}                  & {\color[HTML]{3531FF} $B$}      & {\color[HTML]{3531FF} (Quantum, Classical)} & \multirow{-3}{*}{{\color[HTML]{3531FF} \begin{tabular}[c]{@{}c@{}}Failure\\ probability\end{tabular}}} \\ \hline
{\color[HTML]{036400} KEM-poly32}                                                                        & 21                                                 & 32                                                & 31873                                     & 15                                                         & 12                                                   & 3                                                   & 2                                                 & 2                                                & 4                               & (105, 116)                                  & -113                                                                                                   \\
{\color[HTML]{036400} KEM-poly64}                                                                        & 9                                                  & 64                                                & 7681                                      & 13                                                         & 10                                                   & 3                                                   & 2                                                 & 2                                                & 2                               & (104, 114)                                  & -128                                                                                                   \\
{\color[HTML]{036400} KEM-poly128}                                                                       & 4                                                  & 128                                               & 3329                                      & 12                                                         & 10                                                   & 2                                                   & 2                                                 & 2                                                & 1                               & (101, 111)                                  & -179                                                                                                   \\
{\color[HTML]{036400} Kyber~\cite{kyber_specification}}                                                                             & 2                                                  & 256                                               & 3329                                      & 12                                                         & 10                                                   & 3                                                   & 3                                                 & 2                                                & 1                               & (107, 118)                                  & -139                                                                                                   \\
{\color[HTML]{036400} NewHope~\cite{newhope}}                                                                           & 1                                                  & 512                                               & 12289                                     & 14                                                         & 14                                                   & 2                                                   & 4                                                 & 4                                                & 1                               & (101,  112)                                 & -213                                                                                                   \\ \hline
\end{tabular}
}
\vspace{-15pt}
\end{table}

We target to attain at least $100$-bit post-quantum {core-SVP} security for our lightweight KEM. It will provide equivalent security with AES-$128$~\cite{nist_final_report} and belong to the NIST-level-1 security category. Therefore, the current version of ASCON with $128$ bit security is enough for us. In this section, we discuss the process of finding parameters for Rudraksh.

\begin{figure}[!t]
\vspace{-20pt}
 \begin{minipage}{.45\textwidth}
  \centering
  \includegraphics[width=1\textwidth]{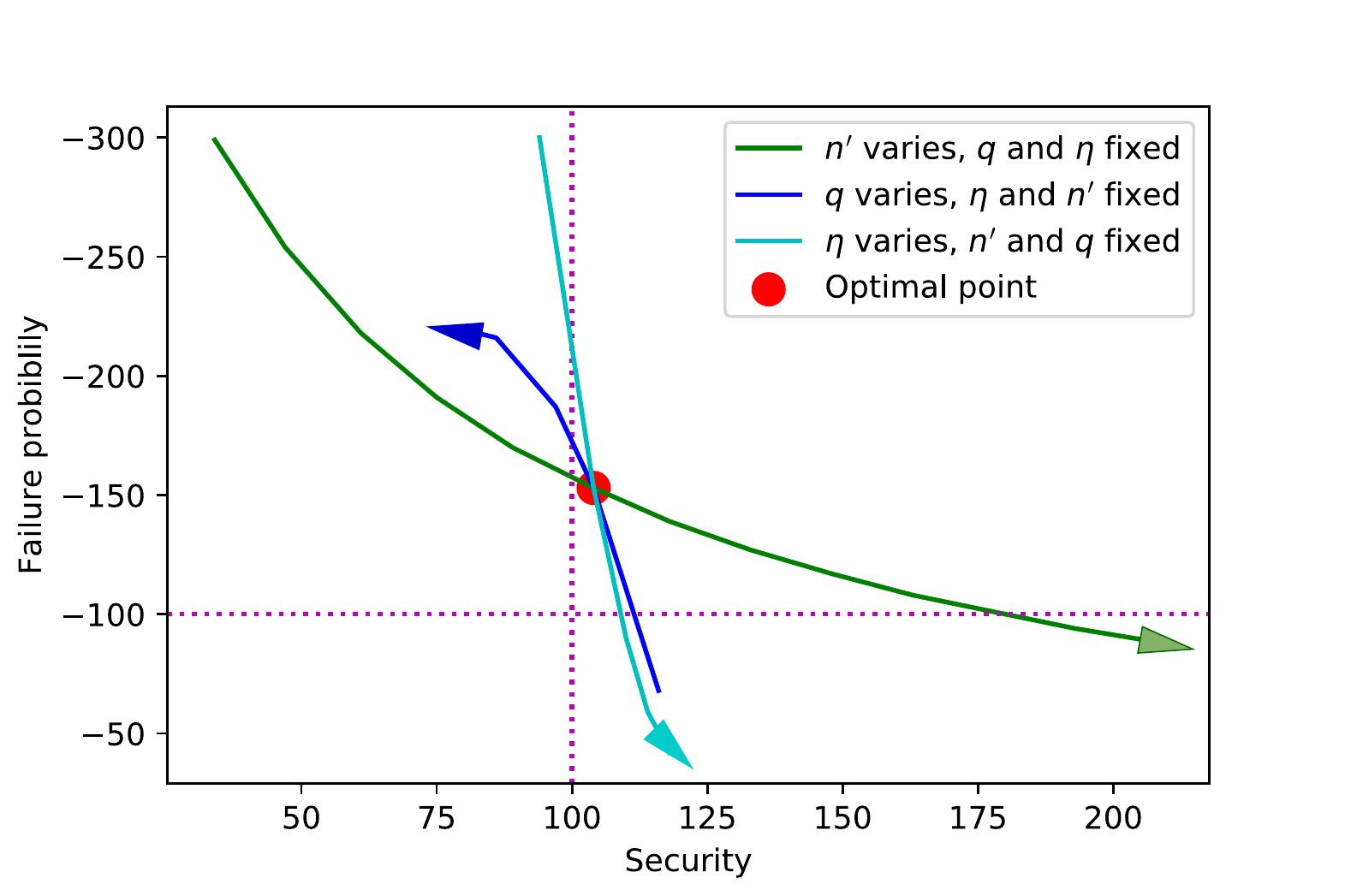}
  \vspace{-10pt}
  \caption{Relation between $n'$, $q$, and $\eta\ (\eta_1/ \eta_2)$ when $n=64$ is fixed (arrows indicate the direction of increase in values). The parameter set of the optimal point is selected for \texttt{KEM-poly64}.}
  \label{fig:parameter-process-64}
\end{minipage}%
\hfill
\begin{minipage}{.5\textwidth}
\vspace{-63pt}
  \centering
  \includegraphics[width=1\textwidth]{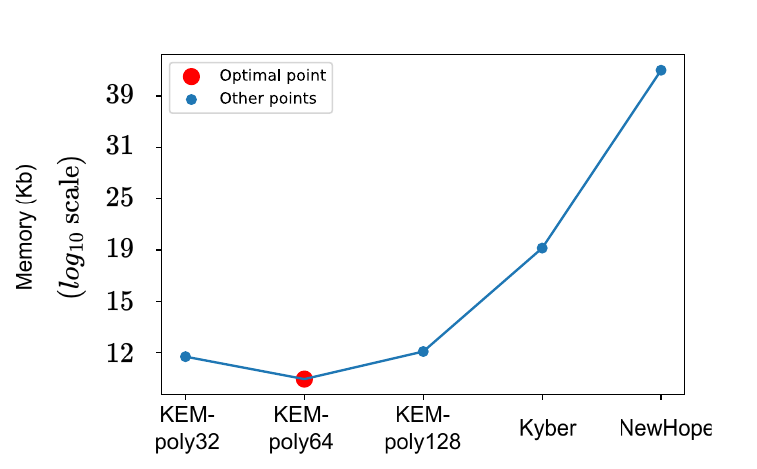}
  \vspace{-10pt}
  \caption{Memory consumption of the KEM depending on the polynomial size}
  \label{fig:64-good-plot}
  \vspace{-25pt}
\end{minipage}
\vspace{-10pt}
\end{figure}

Leaky-LWE estimator~\cite{leaky_estimator} is the state-of-the-art tool to estimate the hardness of the underlying LWE problem. It uses the best-known lattice reduction algorithm Block Korkine-Zolotarev (BKZ)~\cite{BKZ_schnorr, BKZ_nguyen} algorithm. BKZ algorithm primarily estimates the difficulty of solving the shortest vector problem (SVP) in a smaller lattice. This is known as core-SVP hardness. The security of the overall LBC scheme is the hardness of this core-SVP problem with some polynomial overhead. Usually, we ignore this polynomial overhead for a pessimistic estimate of security. The leaky-LWE estimator tool takes the underlying base matrix rank $n'={\ell}\times n$, the modulus $q$, and the standard deviation of secret or error distributions of a scheme as input. It returns both post-quantum and classical bit security of the corresponding scheme. As discussed earlier, while designing a module lattice-based scheme for resource-constrained devices, the two most important parameters are the polynomial-size $n$ and the length of the vector {$\ell$}. 
We have viewed finding the optimal parameter set for our lightweight KEM as a multi-dimensional optimization problem. First, we fixed the polynomial-size $n$ and exhaustively searched all the possible values of other parameters such as the vector length {$\ell$}, modulus $q$, and CBD parameter $\eta_1$,\ $\eta_2$, etc. This is followed by calculating the resource consumption for these parameters.
We have repeated the process for all the power-of-$2$ polynomial sizes to maintain the efficiency of the scheme as it is beneficial for the implementation of several primary building blocks, such as NTT multiplication, $\mathtt{Encode}$, $\mathtt{Decode}$ functions, etc. We also did not explore the polynomial-size below $n = 32$, as the failure probability increases in these cases drastically. 
{We have to increase the modulus $q$ a lot to counteract the high failure probability, affecting the scheme's efficiency. To attain the targeted security, we have to increase the length of vectors, which will also affect the scheme's efficiency.}
We provide optimal parameter sets for three configurations (i) \texttt{KEM-poly32}: with polynomial-size $32$, (ii) \texttt{KEM-poly64}: with polynomial-size $64$, (iii) \texttt{KEM-poly128}: with polynomial-size $128$. The parameters of all these configurations are shown in Tab.~\ref{tab:design_exploration_low}. This table also includes the parameters of Kyber~\cite{kyber_specification}, where the $n=256$, and NewHope~\cite{newhope}, where $n=512=n'$. We also provide the process to find parameters for \texttt{KEM-poly64} in Fig.~\ref{fig:parameter-process-64}. 
{To calculate the failure probabilities, we followed the exhaustive search strategy similar to other LWE-based schemes, such as Kyber, Saber, NewHope, etc. While exploring the KEMs of this work, i.e., KEM-poly32, KEM-poly64, and KEM-poly128, we keep \texttt{repeat}=1 (as $n\leq \texttt{len}_K$).}
%
Memory and area are two primary benchmarks for hardware resource consumption. The memory possesses all the resources used for data usage, which includes all the on-chip memory structures such as Block-RAM (BRAM), distributed RAM, etc. The area contains the configuration logic resources, including look-up tables (LUTs) and logical elements. 
{Now, we discuss the estimated hardware resource usage of all the schemes presented in Tab.~\ref{tab:design_exploration_low} in terms of memory and select the one that can be operated with optimal resources.}
Each polynomial of the secret-key vector $\pmb{s}$ can be generated from $\mathtt{seed}_{\pmb{s}}$. These secret polynomials generate the public-key vector $\pmb{\hat{b}}$ during the key-generation procedure or used in the \texttt{PKE.Dec} (Fig~\ref{fig:kyberpke}, line (3) in \texttt{PKE.KeyGen}) during the decapsulation algorithm. One polynomial length ($n\times \lceil\log_2q\rceil$ bits) memory storage is required for the secret polynomial (for $\pmb{s}$ in \texttt{PKE.KeyGen}, for $\pmb{s'}$ in \texttt{PKE.Enc}), and for the runtime calculation of single polynomial in the public matrix $\pmb{\hat{A}}$. For efficient implementation, another polynomial storage is required for the $n$ roots of unity (or twiddle factors). We use one more polynomial space to save the ciphertext $v$ as well as a vector of polynomial space (${\ell}\times n\times \lceil\log_2q\rceil$ bits) to store the public vector $\pmb{\hat{b}}$ in the key-generation algorithm, and that same space is used to store the ciphertext vector $\pmb{u}$ during \texttt{PKE.Enc}. We also need  $320${-}bit ASCON state register for \texttt{KEM-poly32}, \texttt{KEM-poly64}, \texttt{KEM-poly128}, and $1600${-}bit state register of Keccak for Kyber and NewHope. Extra buffer is often used for post-processing for hash (for $K$ in encapsulation and $K'$, $K''$ in decapsulation) and the pseudorandom number $z$ generated during key generation and used in decapsulation algorithm (for the cases of decryption failure). This buffer size is equivalent to the state register. Therefore, we need memory for four polynomials, one vector of polynomials, states of ASCON or Keccak, and storage for hash output and $z$. We calculate the storage requirement for each of the configurations of Tab.~\ref{tab:design_exploration_low} and present them with the help of Fig.~\ref{fig:64-good-plot}. It is evident from Fig.~\ref{fig:64-good-plot} that \texttt{KEM-poly64} uses the least storage compared to other configurations. Therefore, we select \texttt{KEM-poly64} as our lightweight KEM Rudraksh and present a lightweight (low-resource) hardware implementation.
{The public-key, secret-key, and ciphertext sizes of Rudraksh are 952, 1920, and 760 bytes, respectively. These sizes are equivalent to/slightly higher than those of Kyber, which are 800, 1632, and 768 bytes. This is mainly due to our design decisions and prioritizing resource requirements over communication bandwidth requirements. Usually, lightweight devices use protocols such as BLE and ZigBee for data transmission, which are extremely power- and energy-efficient. Regarding the energy consumption for data transmission, our scheme is similar to that of Kyber. Moreover, our design uses a smaller polynomial size, and the encapsulation/decapsulation module of Rudraksh can start its computation after receiving a public-key/ciphertext polynomial. Hence, we can transfer the key/ciphertext one polynomial at a time without affecting the scheme's efficiency. Although this process does not reduce the total number of byte transfers via a network, because of this interleaved operation between communication and computation, our scheme can operate with a smaller bandwidth. Such techniques need to be explored further in the real-world environment.}
\vspace{-15pt}
\section{Hardware design}\label{sec:hw_design}
 After exploring the theoretical design decisions for developing a lightweight lattice-based KEM, we efficiently implement the scheme and show the scheme's area and latency requirements in hardware. We have chosen Xilinx Virtex-7 {and Artix-7 FPGAs} as our target {devices}. Our hardware design consists of several components. First, we discuss the system architecture; second, we discuss the re-configurable butterfly architecture; and finally, we discuss the datapath of the ASCON-based XOF function. We also discuss the other computational units, such as the CBD sampler, rejection sampling unit, etc. We also discuss our efficient memory organization, as careful memory organization is crucial for memory reduction. Note that reducing memory is important for lightweight design as {most} use cases are memory-constrained IoT devices. They will often share the memory with other systems, and careful memory-reduced design is of utmost importance. Efficient NTT memory access removes the need to reorganize memory, which increases latency overhead. Finally, we describe the scheduling during all the operations.
\subsection{System architecture}\label{sec:system_architecture}
The full system architecture is shown in Fig.~\ref{fig:dataflow}. The seeds of matrix $\pmb{\hat{A}}$, secret $\pmb{s}$, and error $\pmb{e}$ are taken from an external True Random Number Generator for demonstration. 
A controller FSM controls ASCON-XOF and butterfly units synchronously. It enables/gates the CBD/rejection sampler when not required. For example, sampling $\pmb{\hat{A}}$ does not require CBD and hence is gated by the controller. The controller also provides an address offset to avoid memory collision with the public/secret key if we want to store them. 
The ASCON permutation is used for the $\mathcal{H},\ \mathcal{G}$, and $\mathtt{PRF}$ functions. It is the key function to generate the public matrix of polynomials $\pmb{\hat{A}}$. Each coefficient of the polynomials is $\lceil\log_2q\rceil$ {bits} and less than {the} prime {modulus} $q$. Therefore, an additional rejection sampler is needed to discard coefficients $\geq q$, while generating $\pmb{A}$. The ASCON core is also used to generate the pseudo-random number for the CBD sampler module to construct the secret $\pmb{s}$ (or $\pmb{s'}$). The secret is directly stored in 2 NTT memories. The reconfigurable butterfly unit performs the NTT/INTT/point-wise multiplication(PWM) operation. Synchronous memory access for NTT and INTT are ensured in the design. We discuss the details in Sec.~\ref{sec:memory_hw}. PWM is a vector multiplication. {Since} the polynomial-size is small ($64$){,} we use complete NTT multiplication, unlike Kyber, where PWM is complex{,} as it uses incomplete NTT multiplication along with Karatsuba multiplication in the last step as part of PWM.       

\subsection{Reconfigurable butterfly unit}\label{sec:butterfly_hw}
\begin{figure}[!t]
\vspace{-10pt}
\begin{minipage}{.48\linewidth}
  \centering
  \includegraphics[scale=.42]{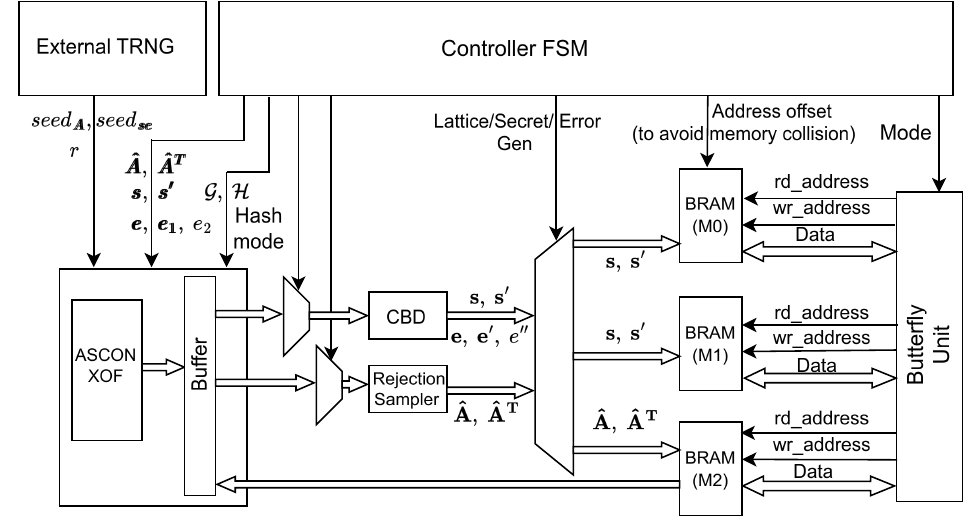}
  \vspace{-10pt}
  \caption{Full system architecture}
  \label{fig:dataflow}
\end{minipage}%
\hfill
\begin{minipage}{.48\linewidth}
  \centering
  \includegraphics[scale=.4]{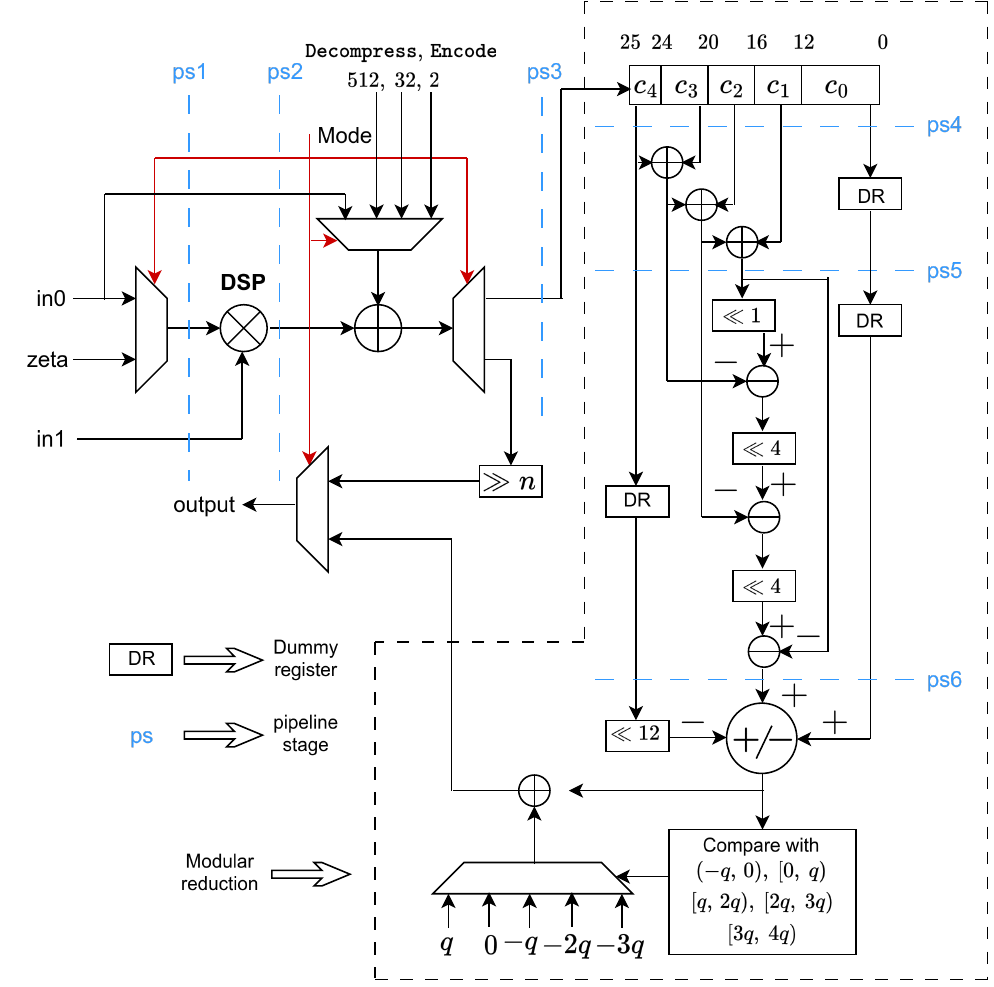}
  \vspace{-10pt}
  \caption{Butterfly module's architecture}
  \label{fig:butterfly}
\end{minipage}%
\vspace{-10pt}
\end{figure}
We introduce a reconfigurable butterfly unit for NTT, INTT, PWM, compress, decompress, encode, and decode functions. The butterfly unit is configured by the 3-bit mode signal provided by the controller (Fig.~\ref{fig:butterfly}). A single DSP unit is used for multiplication. Notably, multiplication is a key operation in all the previous computations and takes a significant area. The DSP unit performs the multiplication {between} the twiddle factor ($\zeta^{j} \bmod{q}$ where $1\leq j\leq n-1$) {and $x_j$} in NTT/INTT operations. 
We adapted Zhang et al.{'s}~\cite{intt_by_2} technique for INTT. Here, the multiplication with $(1/n) \bmod{q}$ of Eq.~\ref{eq:intt} is replaced by the $(1/2)\bmod{q}$ in each butterfly operation. This step eliminated complex multiplication by 1-bit left shift operation at negligible hardware cost.
The butterfly unit also consists of an adder/subtractor, as shown in Fig.~\ref{fig:butterfly}. Finally, it includes a three-stage pipelined shift-and-add modular reduction, ensuring a very low critical path for the design. 
It is important to note that ASCON-XOF is extremely lightweight, and its substitution layer consists of a few 'xor' and 'and' gates. Hence, we use 6-stage pipelines in the butterfly to maintain a low critical path, resulting in a high-frequency design.   

Only the multiplication and the modular reduction are enabled, while the butterfly operates in PWM mode. We employ the butterfly unit to compute $\mathtt{compress}(\pmb{b'},\ 1024)$, defined by $\frac{(\pmb{b'} \ll 10) + {q}/{2}}{q}$ followed by keeping {the} lower 10 bits (similar to~\cite{kyber_code}). The division by $q$ is replaced by multiplication with an approximate value of $1/q$. This procedure includes multiplication with $(2^{32}/q+1)$ followed by $32$ bit right shift. We use a similar technique to compute $\mathtt{compress}({c_m},\ 32)$. Here, we substitute the division by $q$ with multiplication with $(2^{27}/q+1)$ followed by $27$ bit right shift. While computing $\mathtt{Decode}(m)$, we replace the division by $q$ with multiplication with $(2^{30}/q+1)$ and followed by $30$ bit right shift. $\mathtt{decompress}(\pmb{u},\ 1024) = (q\cdot \pmb{u} + 512) \gg 10$ , $\mathtt{decompress}(v,\ 32) = (q\cdot v + 16) \gg 5$, and $\mathtt{encode}(m) = (q\cdot m + 2) \gg 2$ involve a multiplication with $q$. 
All the right shift operations are implemented using a configurable barrel shifter. We describe modular reduction hardware in detail below. 

\textbf{Modular reduction:} 
Modular reduction is one of the crucial parts of the butterfly core. Some of the primarily utilized {modular} reduction algorithms are Montgomery reduction~\cite{Kyber-Kem} and Barrett reduction~\cite{compact_kyber}.
{However, it is important to note that while using Montgomery and Barrett reduction following~\cite{kyber_specification}, one extra multiplication followed by the reduction converts a polynomial to the Montgomery domain. Returning from the Montogomery domain often requires one extra multiplication, costing extra latency for an extra DSP unit.}
Later, for primes like $q=2^m\times k+1$, where $k$ is an odd number, the $k$-reduction~\cite{patrick_longa_ntt} algorithm has been proposed. This reduction algorithm performs better and consumes less area on hardware than the Montgomery or Barrett reduction. Subsequently, the $k^2$-reduction algorithm is introduced by Bisheh-Niasar et al.~\cite{k2-reduction}, which is more efficient than the $k$-reduction algorithm. $k$-reduction algorithm takes input $c$ and outputs $d\equiv k\times c \bmod{q}$, and $k^2$-reduction algorithm takes input $c$ and outputs $d\equiv k^2\times c \bmod{q}$. We can eliminate the extra $k^s$, $s\in \{1,\ 2\}$, by replacing the pre-computed factor $\mathtt{zeta}$ by $k^{-s}\times \mathtt{zeta}$ during NTT or INTT. However, while using the $k^s$-reduction during point-wise multiplication, we have to perform one extra multiplication followed by the reduction to discard the extra $k^s$ factor. It either increases the number of DSPs (containing one multiplication unit) or the latency. However, this extra step is unnecessary if we use the shift-and-add modular reduction technique. 
We used this technique for our prime $q=7681$ and presented it in Alg.~\ref{algo:shift-and-add-reduction}. {It is an improved lightweight hardware-assisted variant of $k$-reduction.} The detailed implementation is shown on the right side of the butterfly unit (Fig.~\ref{sec:butterfly_hw}). The modular reduction only needs addition/subtraction and bit shift operation, making it extremely lightweight and suitable for high-frequency operations. 

\begin{algorithm}[!t]
\caption{Shift-and-add modular reduction {(an improved $k$-reduction~\cite{patrick_longa_ntt})}}
\label{algo:shift-and-add-reduction}
\scriptsize
\Input{$c$ is an integer $\in[0,\ (q-1)^2]$}
\Output{$d\equiv c \bmod{q}$}
$c = c_4||c_3||c_2||c_1||c_0$ \hfill $\rhd |c_0| = 13,\ |c_1| = |c_2| = |c_3| = 4,\ |c_4| = 1$\\
$temp_0 = c_4+c_3$; $temp_1 = temp_0+c_2$; $temp_2 = temp_1+c_1$\\
$temp_3 = (temp_2\ll1)-temp_0$; $temp_4 = (temp_3\ll4)-temp_1$\\
$temp_5 = (temp_4\ll4)-temp_2$; $temp_6 = temp_5+c_0$\\
$res = (-c_4\ll12) + temp_6$\\
\textbf{if} {$((d[15])==1)$} \textbf{then} {$d += q$} \\
\textbf{if} {$(d>q)$} \textbf{then} {$d -= q$} \\
\textbf{if} {$(d>q)$} \textbf{then} {$d -= q$} \\
\textbf{if} {$(d>q)$} \textbf{then} {$d -= q$} \\
\algorithmicreturn{ $d$}
\end{algorithm}
\vspace{-10pt}
\subsection{ASCON core as XOF function }\label{sec:ascon_hw}
\begin{figure}[!b]
\vspace{-10pt}
  \centering
  \includegraphics[scale=.35]{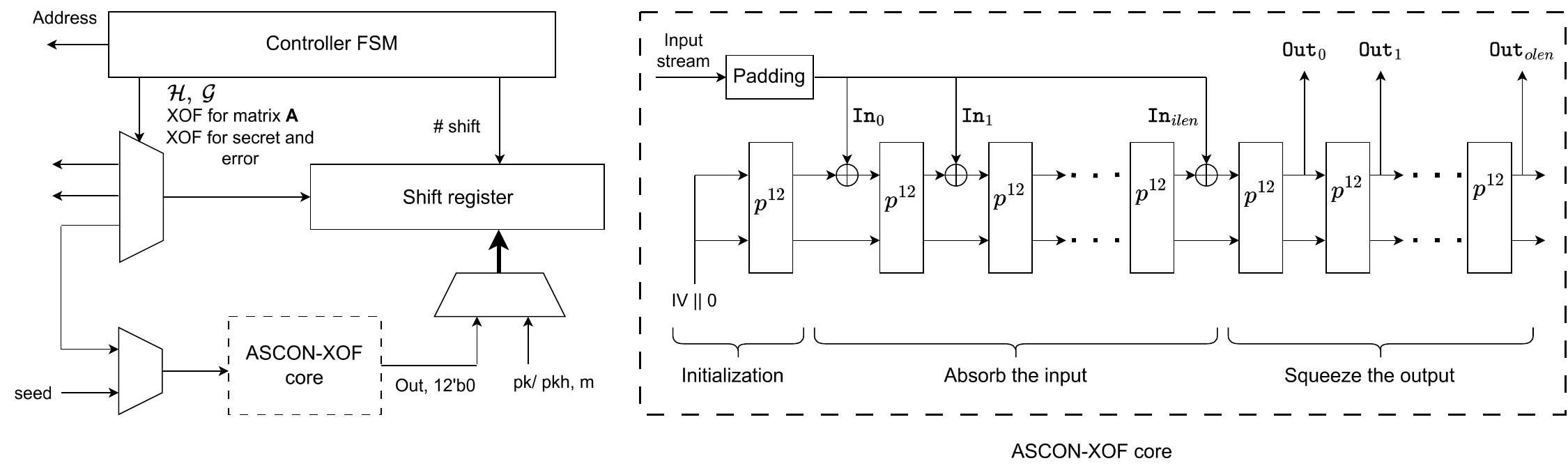}
  \vspace{-10pt}
  \caption{Structure of ASCON XOF hardware}
  \label{fig:ascon}
\end{figure}

ASCON-XOF takes an input of arbitrary size in chunks of 64 bits and generates an output with variable lengths (in chunks of 64 bits). ASCON is also a sponge-based construction like Keccak. It offers a smaller state size of $320$ bits (64-bit rate and 256-bit capacity), whereas the Keccak state register size is $1600$ bits. ASCON-XOF can be implemented with low-area (Fig.~\ref{fig:ascon}). It supports several functional modes which are $\mathtt{PRF}$ to generate the public-matrix $\pmb{\hat{A}}$, the CBD sampler to generate the secret $\pmb{s}$ (or $\pmb{s'}$), error $\pmb{e}$ (or $\pmb{e'}, e''$), and $\mathcal{G},\ \mathcal{H}$. The global controller fixes the mode for this block. 

ASCON-XOF function has three steps: a) initialization, b) absorb, and c) squeeze. The initial state register is pre-computed from $\mathtt{IV}$ in our design to save latency. The second step is to absorb the input stream in the block of $64$ bits. 
In Fig.~\ref{fig:ascon}, $\mathtt{ilen}$ denotes $\lceil\frac{\text{input length} + 1}{64}\rceil$. During absorb, $64$ bits input block is XORed with the first $64$ bits of the state register followed by $p^{12}$. The third step is to squeeze the output bits.  
This process continues until the required length of output is extracted. We denote $\lceil\frac{\text{output length}}{64}\rceil$ by $\mathtt{olen}$ in the figure. 
ASCON permutation $p^{12}$ is the primary building block of the ASCON-XOF function. It is used during all the three steps. This permutation consists of three steps: (i) addition of constant round, (ii) substitution layer, and (iii) linear diffusion layer~\cite{ASCON}. 
 
The rate of input and output block of ASCON is only $64$-bit. ASCON's permutation $p^{12}$ includes only bit-wise XOR, circular shift, and bit-wise AND operations. Therefore, single permutation takes considerably less number of gates than Keccak. Moreover, this implies that the critical path of the ASCON permutation is small and, hence, increases the maximum frequency. The execution of the ASCON permutation at a higher frequency compensates for high clock cycle consumption during absorb and squeeze functions. It helps this design compute with a similar order of latency as Kyber. This design decision assists in attaining an efficient performance with reduced area usage and makes it especially suitable for lightweight designs.  

This ASCON-XOF hardware also contains a $76$-bit buffer/shift register.
This buffer stores the input of the absorb while computing $\mathcal{H}(pk)$. Each coefficient of the vector of polynomials is $13$ bits, and we save the $pk$ in coefficient format one by one. Once a minimum of 64 bits is stored, those bits are used for absorption while new coefficients are introduced in the buffer. The ASCON absorb's input block size is $64$ bits (determined by ASCON rate). We load the first $5$ coefficients of $pk$ ($65$ bits) to the shift register for the first absorb. The input block of the first absorb step consists of the first $4$ {coefficients} and $12$ bits from the {least} significant bits (LSB) of the $5$th {coefficient}. One bit of the $5$th coefficient remains in the shift register. Then, we load the next $5$ coefficients of the $pk$ to the shift register. Now, the shift register holds $66$ bits of input. We use $64$ bits from the LSB (including the remaining $1$ bit of the 1st $5$ coefficients) as the second input block. This process continues until the whole $pk$ is absorbed. As the {greatest common divisor} between $13$ and $64$ is $1$, the minimum size of the shift register needs to be $64+12=76$ to accommodate extra bits of the input stream for all possible cases. 

The same buffer temporarily stores the output blocks when the ASCON block works in PRF mode, producing the public matrix $\pmb{\hat{A}}$. ASCON squeeze generates $64$-bit output after a $12$ round permutation $p^{12}$, and each coefficient size of $\pmb{A}$ is $13$ bits. Then, these coefficients are fed to the rejection sampler. It accepts if the coefficients are less than $q$. So, only four coefficients can be constructed after a single squeeze. There will be $12$ bits remaining in the shift register. After the next squeeze, another $64$ bits output is added to the shift register. To accommodate all the bits, the same $76$-bit shift register is used. From these $76$ bits, $65$ bits from the LSB are utilized to construct the next $5$ coefficients of the matrix $\pmb{\hat{A}}$, and $11$ bits will remain in the shift register. This process will continue until the whole matrix is generated.  

The ASCON block is required to sample the secret $\pmb{s}$ (or $\pmb{s'}$) and error $\pmb{e}$ (or $\pmb{e'}, e''$). The pseudorandom bits created by {ASCON} permutation are fed to the CBD sampler to construct the final secret and error coefficients. Here, the $16$ bytes seed +$1$ byte nonce\footnote{{The nonce is specific to each of the polynomials, and its value depends on its index. For example, this extra byte for ${\pmb{s}_0}$ is $0$, for ${\pmb{s}_8}$ is $8$, for ${\pmb{e}_0}$ is $9$, and for ${\pmb{e}_8}$ is $17$. Two extra bytes are used to generate the polynomial of matrix ${\pmb{\hat{A}}}$; one indicates the column number, and the other is the row number. This method helps to reduce bookkeeping hazards and reduce memory consumption.}} works as the input of the absorb step. Three absorb steps are required as the input block size is $64$ bits. To construct a coefficient of the secret or error, $4$ bits of XOF output are needed. Therefore, $4$-times squeeze is required to generate a single polynomial.
\subsection{Memory organization}\label{sec:memory_hw} 
\begin{figure}[!t]
\vspace{-17pt}
  \centering
  \includegraphics[scale=.5]{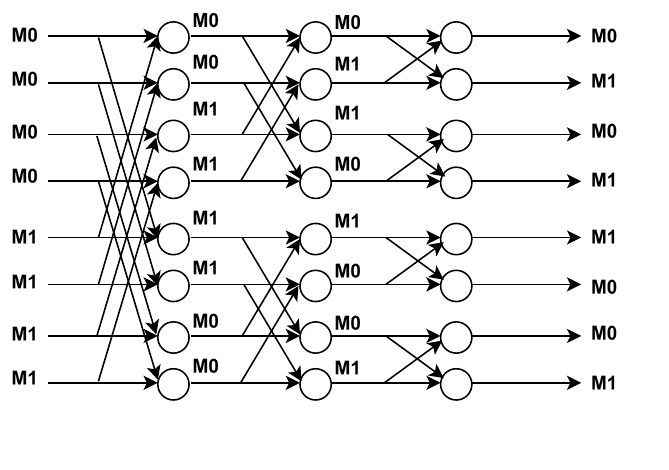}
  \vspace{-25pt}
  \caption{Memory organization of the NTT module}
  \label{fig:NTT_memory}
  \vspace{-15pt}
\end{figure}

One of the key design aspects of our design is to reduce memory as much as possible to make it resource-constraint-device friendly. We took two key approaches to {achieve} this. First, we reduce total memory by carefully choosing {the generation of} lattice $\pmb{\hat{A}}$ and secret $\pmb{s}$. Second, NTT memory organization is done carefully to accommodate {the} minimum BRAM usage for NTT.

We use just two 18K BRAMs for NTT/INTT operations and one 18K BRAM for {run-time} lattice generation and public key storage. While we {generate} $\pmb{\hat{A}}$, the careful design choice of 64-point polynomials gives us the perfect opportunity to synchronize ASCON-XOF-Based $\pmb{\hat{A}}$ generation and $\mathtt{NTT}(\pmb{s})$ operation. For example, generation of $\pmb{A}$ ($13{\cdot}64$ bit) consumes 192 (= $3{\cdot}12$ for absorb + $13{\cdot}12$ for squeeze) cycles. However, we often need to squeeze more to accommodate more coefficients, as some are rejected. This takes 12-24 cycles more on average. Our NTT is a single butterfly design; hence, $\mathtt{NTT}(\pmb{s})$ takes $32{\cdot}6$ = 192 cycles. This careful design ensures that both hardware components work synchronously. This {also} gives us the perfect opportunity for a runtime secret generation, which may not be preferred for NIST standard Kyber as NTT takes significantly more cycles for Kyber. As we need to generate lattice $\pmb{A}$, $\mathtt{NTT}(\pmb{s})$ can be done within that time. This ensures that we can store the secret seeds and generate them every time, reducing the memory requirement for the secret by a significant amount. For example, we need to keep storage of 1 polynomial generation related to the secret generation contrary to 1 vector of a polynomial of the most optimized {Kyber} design~\cite{ml-kem}. If we choose 128-/256-point NTT, $\mathtt{NTT}(\pmb{s})$ takes more cycles, causing run time secret generation to be infeasible.
We also use trivial ML-KEM optimization~\cite{ml-kem}, such as run-time $\pmb{A}$ generation. We have a separate memory for the public key. However, that is not necessary if it is integrated with IoT devices. IoT devices often have extra memory, which can be utilized to communicate with another party.

NTT memory is implemented with 2 separate memory as shown in Fig.~\ref{fig:NTT_memory}. Once the secret is generated, 1st half is written in one BRAM, say M0, whereas 2nd half is written in M1 as coeff[0], and coeff[32] is required in the first stage. At every level, writing is swapped to ensure the next stage data is available from 2 different memory. This strategy ensures streamlined dataflow even with a single port write-enabled memory. 
We are using 3, 18K BRAMs for this implementation. However, total memory is not used. For example, M0 \& M1 need just $32{\cdot}13$-bit memory each (416 bit, 2.3\% of 18K memory).   

\subsection{Scheduling}\label{sec:scheduling_hw}
\begin{figure}[!b]
  \vspace{-10pt}
  \hspace*{-20pt}
  \includegraphics[scale=.5]{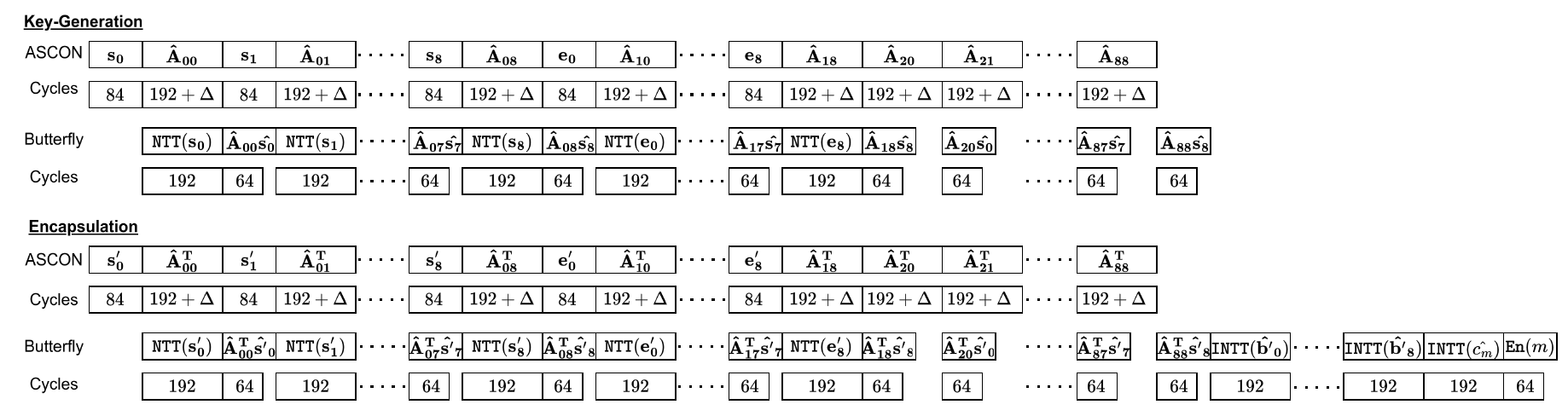}
  \vspace{-20pt}
  \caption{Scheduling with ASCON and butterfly module }
  \label{fig:scheduling}
\end{figure}
Scheduling is an important aspect of KEM hardware design. There are two parallel components of the KEM data path: (i) butterfly unit, which computes NTT, INTT, and PWM as well as encode, compress, and decompress, and (ii) Hash/PRF functions (ASCON-XOF for our case, SHA3 and SHAKE for Kyber). These two components are independent. This allows us to schedule synchronously, as shown in Fig.~\ref{fig:scheduling}. Note that both components are specially required for the keygen and encrypt phases. Decrypt only needs butterfly, whereas FO-related functions require ASCON-XOF only. We sample using ASCON-XOF and CBD, which needs 84 cycles; then, one polynomial is sampled using rejection sampling followed by ASCON XOF. It is important to note that rejection sampling, in this case, takes at least 192 cycles, which is enough to calculate 64-point NTT. Then, while sampling the next secret, we can multiply and accumulate it as that takes fewer cycles. A lightweight ASCON core takes multiple cycles to create {the matrix $\pmb{\hat{A}}$ or $\pmb{\hat{A}^T}$}, even if the secret is stored. We have the option to store the entire secret in memory. However, runtime generation of the secret polynomial costs only 160 cycles of latency in the key generation/encryption function, which is negligible. As we are targeting lightweight design for energy and area-constrained IoT devices, we have taken this approach to reduce memory further.  

\subsection{Other computational units}\label{sec:other_hw}
We require two more components: rejection sampling and CBD sampler. Rejection sampling is a part of matrix $\pmb{\hat{A}}$ or $\pmb{\hat{A}^T}$ generation. It checks and accepts if the ASCON-XOF generated $13$ bits output is less than $q$. Otherwise, it rejects those $13$ bits and proceeds with the next $13$ bits. The implementation of this component is not constant time. However, it does not affect the security as the matrix $\pmb{\hat{A}}$ is a public matrix. 
The CBD sampler is used to sample coefficients of the secret polynomials $\pmb{s}$, $\pmb{s'}$, and the error polynomials $\pmb{e}$, $\pmb{e'}$, $e''$. Each coefficient of these polynomials is constructed from $4$ bits output of the ASCON-XOF. These $4$ bits output can be denoted as a[0:3]. The coefficient is implemented by the following operation $b = \mathtt{HW}(a[0:1])-\mathtt{HW}(a[2:3])$. Then the coefficient value $b\in[-2,\ 2]$. In other words, secret/error is sampled by calculating the Hamming {distance} of two 2-bit numbers.

\vspace{-10pt}
\section{Results}\label{sec:results}
In this section, we will discuss the implementation results and compare them with the state-of-the-art KEM designs.
\vspace{-10pt}
\subsection{Resource consumption of submodules} 
\begin{table}[!b]
\vspace{-10pt}
\begin{minipage}{.45\linewidth}
\centering
\caption{Resource requirements of various modular reductions}
\label{tab:modular_reduction}
\resizebox{1\columnwidth}{!}{%
\begin{tabular}{cccc}
\hline
\multicolumn{1}{c|}{{\color[HTML]{3531FF} }}                                    & \multicolumn{3}{c}{{\color[HTML]{3531FF} Area}}                                   \\ \cline{2-4} 
\multicolumn{1}{c|}{\multirow{-2}{*}{{\color[HTML]{3531FF} Modular reduction}}} & {\color[HTML]{3531FF} } & {\color[HTML]{3531FF} DSP} & {\color[HTML]{3531FF} LUT} \\ \hline
\multicolumn{1}{c|}{{\color[HTML]{036400} \textbf{Shift-and-add}}}              & \textbf{}               & \textbf{0}                 & \textbf{102}               \\
\multicolumn{1}{c|}{{\color[HTML]{036400} Montgomery+Barrett$^*$}~\cite{kyber_specification}}              &                         & 2                          & 13                         \\
\multicolumn{1}{c|}{{\color[HTML]{036400} $k^2$-reduction$^*$}~\cite{Ayesha-k2-reduction}}                 &                         & 0                          & 132                        \\
\multicolumn{1}{c|}{{\color[HTML]{036400} $k^2$-reduction$^*$}~\cite{k2-reduction}}                 &                         & 0                          & 80                         \\ \hline
\multicolumn{4}{l}{{\color[HTML]{333333} $^*$ one extra polynomial multiplication is required}}                                                                     \\ 
\end{tabular}
}
\end{minipage}%
\hfill
\begin{minipage}{.5\linewidth}
\centering
\caption{Submodules area requirements}
\label{tab:area_submodule}
\resizebox{1\columnwidth}{!}{%
\begin{tabular}{ll|cccc}
\hline
\multicolumn{2}{c|}{{\color[HTML]{3531FF} }} & \multicolumn{4}{c}{{\color[HTML]{3531FF} Area (client/server)}} \\ \cline{3-6} 
\multicolumn{2}{c|}{\multirow{-2}{*}{{\color[HTML]{3531FF} Modular reduction}}} & {\color[HTML]{3531FF} LUT} & {\color[HTML]{3531FF} FF} & {\color[HTML]{3531FF} BRAM} & {\color[HTML]{3531FF} DSP} \\ \hline
\multicolumn{2}{l|}{Butterfly} & 514/514 & 325/325 & 0/0 & 1/1 \\
\multicolumn{1}{l|}{{\color[HTML]{036400} }} & Reduction & 102/102 & 27/27 & 0/0 & 0/0 \\
\multicolumn{2}{l|}{ASCON-XOF} & 689/689 & 326/326 & 0/0 & 0/0 \\
\multicolumn{1}{l|}{{\color[HTML]{036400} }} & Permutation $p^{12}$ & 685/685 & 321/321 & 0/0 & 0/0 \\
\multicolumn{1}{l|}{} & Round counter & 4/4 & 5/5 & 0/0 & 0/0 \\
\multicolumn{2}{l|}{CBD(x2)} & 10/10 & 0/0 & 0/0 & 0/0 \\
\multicolumn{2}{l|}{Rejection sampling} & 17/17 & 12/12 & 0/0 & 0/0 \\
\multicolumn{2}{l|}{} &  &  &  &  \\
\multicolumn{2}{l|}{} &  &  &  &  \\
\multicolumn{2}{l|}{} &  &  &  &  \\
\multicolumn{2}{l|}{\multirow{-4}{*}{\begin{tabular}[c]{@{}l@{}}Top logic \\ (ASCON buffer\\ +FSM controller\\ +verify)\end{tabular}}} & \multirow{-4}{*}{1583/1639} & \multirow{-4}{*}{831/750} & \multirow{-4}{*}{1.5/1.5} & \multirow{-4}{*}{0/0} \\ \hline
\multicolumn{2}{l|}{\textbf{Total}} & 2813/2869 & 1494/1413 & 1.5/1.5 & 1/1 \\ \hline
\end{tabular}
}
\end{minipage}
\vspace{-10pt}
\end{table} 

Resource consumption for each component, followed by full hardware, is presented in Tab.~\ref{tab:area_submodule}. Our butterfly design requires multiplication, which is realized by a DSP unit. The reconfigurable butterfly consumes only 514 LUT and 325 Flip-flops in addition to 1 DSP unit. Modular reduction is one of the key components of the butterfly unit. 
{We have explored multiple strategies for $q=7681$ and shown in Tab.~\ref{tab:modular_reduction}. Although $k^2$-reduction~\cite{k2-reduction} requires the least area, one extra multiplication followed by the reduction is performed during PWM to discard the extra $k^2$ factor, as discussed in Sec.~\ref{sec:butterfly_hw}. Therefore, we conclude that the shift-and-add modular reduction is the least hardware-intensive.}

{Another key component of the datapath is {the} ASCON-XOF hardware. Permutation consumes maximum area with 685 LUT and 321 flip-flops.} The top includes an FSM controller, a 76-bit buffer/shift register for ASCON-XOF and verify logic and 3-block RAMs. Overall, the controller is the key contributor in terms of area. This exploration indicates that in case of low latency requirement, an HW-SW codesign approach can also be taken to minimize the area further. We have used three 18K BRAMs in total. 18K BRAMs are considered 0.5 BRAM in FPGA architecture. 2 BRAMs (M0, M1) are used for NTT/INTT operations, and another BRAM (M2) has been used for public key storage. However, we do not use the entire memory. For M0, M1, only 2.3\% of the BRAM has been used, whereas M2 uses $\sim40\%$ of the BRAM. {Overall, our implementation of Rudraksh consumes only $2813$/$2869$ LUT and $1494$/$1413$ Flip-flops with a single DSP.}

{Using ASCON-XOF instead of Keccak comes with a drawback; it requires more clock cycles to generate the same amount of pseudo-random numbers due to its lightweight state registers. However, the highly lightweight datapath enables higher-frequency operations, and our 6-stage pipelined architecture of the butterfly module keeps the critical path low. The server runs the key-generation and decapsulation algorithm, and the client, which runs only the encapsulation algorithm, can operate at 431MHz and 400MHz, respectively. Therefore, although the key generation, encapsulation, and decapsulation {use} 23310, 28114, and 35110 {cycles}, respectively, the execution times are 54 $\mu$s, 70 $\mu$s, and 81 $\mu$s, respectively. Latency is often not the utmost priority in resource/energy-constraint IoT devices, though latency overhead is reasonable due to the careful design of the datapath.}

{We compare the area cost and execution time of Rudraksh (which uses ASCON), and Kyber (which uses Keccak) in Tab.~\ref{tab:ascon-benefit} to exhibit the advantages of ASCON and other design decisions in Rudraksh. This table uses the equivalent number of slices (ENS), where we convert all FPGA components to an equivalent gate model to demonstrate overall area consumption~\cite{Ayesha-k2-reduction}. In Rudraksh, ASCON consumes 4.3$\times$ less LUT and 4.9$\times$ less FF than the Keccak in Kyber~\cite{compact_kyber}. ASCON takes 0.19 $\mu$s to generate a secret polynomial (64$\times$4 bits), whereas Keccak takes 0.49 $\mu$s for a secret polynomial (256$\times$6 bits) in Kyber~\cite{compact_kyber}. Rudraksh takes the least area with a competitive time due to our low critical path. Our complete architecture except ASCON takes 2$\times$ less LUT, 2.6$\times$ less FF, 2$\times$ less BRAM, and 2$\times$ less DSP compared to Kyber~\cite{compact_kyber} architecture except Keccak. Finally, in terms of ENS, the ASCON in Rudraksh consumes 4.3$\times$ fewer ENS compared to the Keccak in Kyber~\cite{compact_kyber} and 8.5$\times$ fewer ENS compared to the Keccak in Kyber~\cite{HE2024167}.}

\begin{table}[!t]
\vspace{-5pt}
\centering
\caption{{Benefits of ASCON in Rudraksh compared to Keccak in Kyber}}
\vspace{-5pt}
\label{tab:ascon-benefit}
\resizebox{1\textwidth}{!}{%
\begin{tabular}{llcccccccc}
\hline
\multicolumn{2}{c|}{{\color[HTML]{3531FF} }} & \multicolumn{5}{c|}{{\color[HTML]{3531FF} Area (client/server)}} & \multicolumn{1}{c|}{{\color[HTML]{3531FF} }} & \multicolumn{1}{c|}{{\color[HTML]{3531FF} }} & {\color[HTML]{3531FF} } \\ \cline{3-7}
\multicolumn{2}{c|}{\multirow{-2}{*}{{\color[HTML]{3531FF} Scheme}}} & {\color[HTML]{3531FF} LUT} & {\color[HTML]{3531FF} FF} & {\color[HTML]{3531FF} Slice} & {\color[HTML]{3531FF} BRAM} & \multicolumn{1}{c|}{{\color[HTML]{3531FF} DSP}} & \multicolumn{1}{c|}{\multirow{-2}{*}{{\color[HTML]{3531FF} \begin{tabular}[c]{@{}c@{}}ENS$^{**}$\\ (client/server)\end{tabular}}}} & \multicolumn{1}{c|}{\multirow{-2}{*}{{\color[HTML]{3531FF} \begin{tabular}[c]{@{}c@{}}Freq.\\ (MHz)\end{tabular}}}} & \multirow{-2}{*}{{\color[HTML]{3531FF} \begin{tabular}[c]{@{}c@{}}Time to generate a\\ secret polynomial ($\mu$s)\end{tabular}}} \\ \hline
\multicolumn{2}{l|}{{\color[HTML]{036400} \textbf{Rudraksh}}} & \textbf{2813/2869} & \textbf{1494/1413} & \textbf{0} & \textbf{1.5} & \multicolumn{1}{c|}{\textbf{1}} & \multicolumn{1}{c|}{\textbf{1098/1112}} & \multicolumn{1}{c|}{} &  \\
\multicolumn{1}{l|}{{\color[HTML]{036400} \textbf{}}} & \multicolumn{1}{l|}{{\color[HTML]{036400} \textbf{ASCON}}} & \textbf{689} & \textbf{326} & \textbf{0} & \textbf{0} & \multicolumn{1}{c|}{\textbf{0}} & \multicolumn{1}{c|}{\textbf{173}} & \multicolumn{1}{c|}{\multirow{-2}{*}{\textbf{400/431}}} & \multirow{-2}{*}{\textbf{\begin{tabular}[c]{@{}c@{}}0.19\\ (64$\times$4 bits)\end{tabular}}} \\ \hline
\multicolumn{2}{l|}{{\color[HTML]{036400} Kyber\cite{HE2024167}}} & 4777/4993 & 2661/2765 & 1395/1452 & 2.5 & \multicolumn{1}{c|}{0} & \multicolumn{1}{c|}{3080/3191} & \multicolumn{1}{c|}{} &  \\
\multicolumn{1}{l|}{{\color[HTML]{036400} }} & \multicolumn{1}{l|}{{\color[HTML]{036400} Keccak}} & 2826 & 1629 & 770 & 0 & \multicolumn{1}{c|}{0} & \multicolumn{1}{c|}{1477} & \multicolumn{1}{c|}{\multirow{-2}{*}{244}} & \multirow{-2}{*}{-} \\ \hline
\multicolumn{2}{l|}{{\color[HTML]{036400} Kyber\cite{compact_kyber}}} & 6785/7412 & 3981/4644 & 1899/2126 & 3 & \multicolumn{1}{c|}{2} & \multicolumn{1}{c|}{4384/4767} & \multicolumn{1}{c|}{} &  \\
\multicolumn{1}{l|}{{\color[HTML]{036400} }} & \multicolumn{1}{l|}{{\color[HTML]{036400} Keccak}} & 2966/2956 & 1610/1610 & - & 0 & \multicolumn{1}{c|}{0} & \multicolumn{1}{c|}{742/739} & \multicolumn{1}{c|}{\multirow{-2}{*}{161}} & \multirow{-2}{*}{\begin{tabular}[c]{@{}c@{}}0.49\\ (256$\times$6 bits)\end{tabular}} \\ \hline
\multicolumn{10}{l}{$^{**}$ENS (equivalent number of slices) = Slice $+$ DSP$\times100$ $+$ BRAM$\times196$ $+$ LUT$/4$}
\end{tabular}
}
\end{table}

\vspace{-10pt}
\subsection{Comparison with the state-of-the-art}

\begin{table}[btp]
\caption{{Comparison of implementation of Rudraksh (KEM-poly64) with the state-of-the-art schemes. Freq. represents frequency, Exec time represents execution time, and {k} denotes 1000x. KG, Enc, and Dec represent key-generation, encapsulation, and decapsulation, respectively. All the KEMs, except the $*$ and $\dag$ ones, belong to the NIST-level-1 category.}}
\label{tab:sota}
\resizebox{1\textwidth}{!}{%
\begin{tabular}{lcccllllcccc}
\hline
\multicolumn{1}{c|}{{\color[HTML]{3531FF} }} & \multicolumn{1}{c|}{{\color[HTML]{3531FF} }} & \multicolumn{1}{c|}{{\color[HTML]{3531FF} }} & \multicolumn{5}{c|}{{\color[HTML]{3531FF} Area}} & \multicolumn{1}{c|}{{\color[HTML]{3531FF} }} & \multicolumn{1}{c|}{{\color[HTML]{3531FF} Freq.}} & \multicolumn{1}{c|}{{\color[HTML]{3531FF} Exec time($\mu$s)}} & {\color[HTML]{3531FF} T$\times$A} \\ \cline{4-8} \cline{11-11}
\multicolumn{1}{c|}{\multirow{-2}{*}{{\color[HTML]{3531FF} Scheme}}} & \multicolumn{1}{c|}{\multirow{-2}{*}{{\color[HTML]{3531FF} Platform}}} & \multicolumn{1}{c|}{\multirow{-2}{*}{{\color[HTML]{3531FF} \begin{tabular}[c]{@{}c@{}}\small{client}/\\ \small{server}\end{tabular}}}} & \multicolumn{5}{c|}{{\color[HTML]{3531FF} LUT/FF/Slice/BRAM/DSP}} & \multicolumn{1}{c|}{\multirow{-2}{*}{{\color[HTML]{3531FF} ENS$^{**}$}}} & \multicolumn{1}{c|}{{\color[HTML]{3531FF} (MHz)}} & \multicolumn{1}{c|}{{\color[HTML]{3531FF} KG/Enc/Dec}} & {\color[HTML]{3531FF} (ENS$\times$ms)} \\ \hline
\multicolumn{1}{l|}{{\color[HTML]{036400} }} & \multicolumn{1}{c|}{} & \multicolumn{1}{c|}{\small{client}} & \multicolumn{5}{c|}{\textbf{2813/1494/0/1.5/1}} & \multicolumn{1}{c|}{\textbf{1098}} & \multicolumn{1}{c|}{\textbf{400}} & \multicolumn{1}{c|}{} &  \\
\multicolumn{1}{l|}{{\color[HTML]{036400} }} & \multicolumn{1}{c|}{\multirow{-2}{*}{\textbf{Virtex-7}}} & \multicolumn{1}{c|}{\small{server}} & \multicolumn{5}{c|}{\textbf{2869/1413/0/1.5/1}} & \multicolumn{1}{c|}{\textbf{1112}} & \multicolumn{1}{c|}{\textbf{431}} & \multicolumn{1}{c|}{\multirow{-2}{*}{\textbf{54/70/81}}} & \multirow{-2}{*}{\textbf{60/77/90}} \\ \cline{2-12} 
\multicolumn{1}{l|}{{\color[HTML]{036400} }} & \multicolumn{1}{c|}{} & \multicolumn{1}{c|}{\small{client}} & \multicolumn{5}{c|}{\textbf{2776/1487/0/1.5/1}} & \multicolumn{1}{c|}{\textbf{1088}} & \multicolumn{1}{c|}{\textbf{387}} & \multicolumn{1}{c|}{} &  \\
\multicolumn{1}{l|}{\multirow{-4}{*}{{\color[HTML]{036400} \textbf{\begin{tabular}[c]{@{}l@{}}Rudraksh \\ (KEM-poly64)\end{tabular}}}}} & \multicolumn{1}{c|}{\multirow{-2}{*}{\textbf{Artix-7}}} & \multicolumn{1}{c|}{\small{server}} & \multicolumn{5}{c|}{\textbf{2839/1413/0/1.5/1}} & \multicolumn{1}{c|}{\textbf{1104}} & \multicolumn{1}{c|}{\textbf{367}} & \multicolumn{1}{c|}{\multirow{-2}{*}{\textbf{64/73/96}}} & \multirow{-2}{*}{\textbf{71/79/106}} \\ \hline
\multicolumn{1}{l|}{{\color[HTML]{036400} }} & \multicolumn{1}{c|}{} & \multicolumn{1}{c|}{\small{client}} & \multicolumn{5}{c|}{4777/2661/1395/2.5/0} & \multicolumn{1}{c|}{3080} & \multicolumn{1}{c|}{} & \multicolumn{1}{c|}{} &  \\
\multicolumn{1}{l|}{\multirow{-2}{*}{{\color[HTML]{036400} Kyber\cite{HE2024167}}}} & \multicolumn{1}{c|}{\multirow{-2}{*}{Kintex-7}} & \multicolumn{1}{c|}{\small{server}} & \multicolumn{5}{c|}{4993/2765/1452/2.5/0} & \multicolumn{1}{c|}{3191} & \multicolumn{1}{c|}{\multirow{-2}{*}{244}} & \multicolumn{1}{c|}{\multirow{-2}{*}{278/416/552}} & \multirow{-2}{*}{887/1281/1761} \\ \hline
\multicolumn{1}{l|}{{\color[HTML]{036400} Kyber\cite{9926344}}} & \multicolumn{1}{c|}{Artix-7} & \multicolumn{1}{c|}{} & \multicolumn{5}{c|}{8966/9173/3186/10.5/6} & \multicolumn{1}{c|}{8086} & \multicolumn{1}{c|}{204} & \multicolumn{1}{c|}{11.5/17.3/23.5} & 93/140/190 \\ \hline
\multicolumn{1}{l|}{{\color[HTML]{036400} }} & \multicolumn{1}{c|}{} & \multicolumn{1}{c|}{\small{client}} & \multicolumn{5}{c|}{6785/3981/1899/3/2} & \multicolumn{1}{c|}{4384} & \multicolumn{1}{c|}{} & \multicolumn{1}{c|}{} &  \\
\multicolumn{1}{l|}{\multirow{-2}{*}{{\color[HTML]{036400} Kyber\cite{compact_kyber}}}} & \multicolumn{1}{c|}{\multirow{-2}{*}{Artix-7}} & \multicolumn{1}{c|}{\small{server}} & \multicolumn{5}{c|}{7412/4644/2126/3/2} & \multicolumn{1}{c|}{4767} & \multicolumn{1}{c|}{\multirow{-2}{*}{161}} & \multicolumn{1}{c|}{\multirow{-2}{*}{23.4/30.5/41.3}} & \multirow{-2}{*}{112/134/197} \\ \hline
\multicolumn{1}{l|}{{\color[HTML]{036400} }} & \multicolumn{1}{c|}{Virtex-7} & \multicolumn{1}{c|}{} & \multicolumn{5}{c|}{13745/11107/4590/14/8} & \multicolumn{1}{c|}{11571} & \multicolumn{1}{c|}{245} & \multicolumn{1}{c|}{8.8/12.2/17.9} & 102/141/207 \\ \cline{2-12} 
\multicolumn{1}{l|}{\multirow{-2}{*}{{\color[HTML]{036400} Kyber\cite{DBLP:journals/iacr/DangFAMNG20}}}} & \multicolumn{1}{c|}{Artix-7} & \multicolumn{1}{c|}{} & \multicolumn{5}{c|}{11864/10348/3989/15/8} & \multicolumn{1}{c|}{10695} & \multicolumn{1}{c|}{210} & \multicolumn{1}{c|}{-/14.3/20.9} & -/153/224 \\ \hline
\multicolumn{1}{l|}{{\color[HTML]{036400} Kyber\cite{DBLP:journals/iacr/BanerjeeUC19}}} & \multicolumn{1}{c|}{Artix-7} & \multicolumn{1}{c|}{} & \multicolumn{5}{c|}{14975/2539/4173/14/11} & \multicolumn{1}{c|}{11761} & \multicolumn{1}{c|}{25} & \multicolumn{1}{c|}{2980/5268/5692} & 35{k}/62{k}/67{k} \\ \hline
\multicolumn{1}{l|}{{\color[HTML]{036400} }} & \multicolumn{1}{c|}{} & \multicolumn{1}{c|}{\small{client}} & \multicolumn{5}{c|}{6745/3528/1855/1/11} & \multicolumn{1}{c|}{4838} & \multicolumn{1}{c|}{167} & \multicolumn{1}{c|}{} &  \\
\multicolumn{1}{l|}{\multirow{-2}{*}{{\color[HTML]{036400} Frodo\cite{frodo_hardware}}}} & \multicolumn{1}{c|}{\multirow{-2}{*}{Artix-7}} & \multicolumn{1}{c|}{\small{server}} & \multicolumn{5}{c|}{7220/3549/1992/1/16} & \multicolumn{1}{c|}{5593} & \multicolumn{1}{c|}{162} & \multicolumn{1}{c|}{\multirow{-2}{*}{20{k}/20{k}/21{k}}} & \multirow{-2}{*}{110{k}/96{k}/116{k}} \\ \hline
\multicolumn{1}{l|}{{\color[HTML]{036400} NewHope\cite{intt_by_2}}} & \multicolumn{1}{c|}{Artix-7} & \multicolumn{1}{c|}{} & \multicolumn{5}{c|}{6780/4026/-/7/2} & \multicolumn{1}{c|}{3267} & \multicolumn{1}{c|}{200} & \multicolumn{1}{c|}{21/33/12.5} & 69/108/41 \\ \hline
\multicolumn{1}{l|}{{\color[HTML]{036400} LightSaber\cite{DBLP:journals/tches/RoyB20}}} & \multicolumn{1}{c|}{} & \multicolumn{1}{c|}{} & \multicolumn{5}{c|}{23686/9805/0/2/0} & \multicolumn{1}{c|}{6314} & \multicolumn{1}{c|}{150} & \multicolumn{1}{c|}{18.4/26.9/33.6} & 116/170/212 \\ \cline{1-1} \cline{3-12} 
\multicolumn{1}{l|}{{\color[HTML]{036400} Espada$^\dagger$\cite{scabbard-tecs}}} & \multicolumn{1}{c|}{} & \multicolumn{1}{c|}{} & \multicolumn{5}{c|}{18741/18823/-/14/48} & \multicolumn{1}{c|}{12229} & \multicolumn{1}{c|}{250} & \multicolumn{1}{c|}{92.2/154.3/219.3} & 1128/1887/2682 \\ \cline{1-1} \cline{3-12} 
\multicolumn{1}{l|}{{\color[HTML]{036400} Sable$^\dagger$\cite{scabbard-tecs}}} & \multicolumn{1}{c|}{} & \multicolumn{1}{c|}{} & \multicolumn{5}{c|}{17092/11280/-/2/0} & \multicolumn{1}{c|}{4665} & \multicolumn{1}{c|}{250} & \multicolumn{1}{c|}{18.9/23.6/29.0} & 88/110/135 \\ \cline{1-1} \cline{3-12} 
\multicolumn{1}{l|}{{\color[HTML]{036400} Florete$^\dagger$\cite{scabbard-tecs}}} & \multicolumn{1}{c|}{\multirow{-4}{*}{\begin{tabular}[c]{@{}c@{}}Zynq\\ Ultra-\\ scale+\end{tabular}}} & \multicolumn{1}{c|}{} & \multicolumn{5}{c|}{28281/16029/-/2/140} & \multicolumn{1}{c|}{21462} & \multicolumn{1}{c|}{250} & \multicolumn{1}{c|}{28.3/56.7/84.4} & 607/1217/1811 \\ \hline
\multicolumn{1}{l|}{{\color[HTML]{036400} }} & \multicolumn{1}{c|}{} & \multicolumn{1}{c|}{\small{KG}} & \multicolumn{5}{c|}{49001/39957/9357/2.5/45} & \multicolumn{1}{c|}{26598} & \multicolumn{1}{c|}{} & \multicolumn{1}{c|}{} &  \\
\multicolumn{1}{l|}{{\color[HTML]{036400} }} & \multicolumn{1}{c|}{} & \multicolumn{1}{c|}{\small{Enc}} & \multicolumn{5}{c|}{31494/25120/6652/2.5/0} & \multicolumn{1}{c|}{15016} & \multicolumn{1}{c|}{} & \multicolumn{1}{c|}{} &  \\
\multicolumn{1}{l|}{\multirow{-3}{*}{{\color[HTML]{036400} \begin{tabular}[c]{@{}l@{}}NTRU-HRSS701\\ \cite{DBLP:journals/tc/DangMG23}\end{tabular}}}} & \multicolumn{1}{c|}{} & \multicolumn{1}{c|}{\small{Dec}} & \multicolumn{5}{c|}{37702/34441/8032/2.5/45} & \multicolumn{1}{c|}{22448} & \multicolumn{1}{c|}{\multirow{-3}{*}{300}} & \multicolumn{1}{c|}{\multirow{-3}{*}{172.7/7.4/29.4}} & \multirow{-3}{*}{4593/111/660} \\ \cline{1-1} \cline{3-12} 
\multicolumn{1}{l|}{{\color[HTML]{036400} }} & \multicolumn{1}{c|}{} & \multicolumn{1}{c|}{\small{KG}} & \multicolumn{5}{c|}{41047/39037/7968/6/45} & \multicolumn{1}{c|}{23906} & \multicolumn{1}{c|}{} & \multicolumn{1}{c|}{} &  \\
\multicolumn{1}{l|}{{\color[HTML]{036400} }} & \multicolumn{1}{c|}{} & \multicolumn{1}{c|}{\small{Enc}} & \multicolumn{5}{c|}{26325/17568/4638/5/0} & \multicolumn{1}{c|}{12200} & \multicolumn{1}{c|}{\multirow{-2}{*}{250}} & \multicolumn{1}{c|}{} &  \\
\multicolumn{1}{l|}{\multirow{-3}{*}{{\color[HTML]{036400} \begin{tabular}[c]{@{}l@{}}NTRU-HPS677\\ \cite{DBLP:journals/tc/DangMG23}\end{tabular}}}} & \multicolumn{1}{c|}{\multirow{-6}{*}{\begin{tabular}[c]{@{}c@{}}Zynq\\ Ultra-\\ scale+\end{tabular}}} & \multicolumn{1}{c|}{\small{Dec}} & \multicolumn{5}{c|}{29935/19511/5217/2.5/45} & \multicolumn{1}{c|}{17691} & \multicolumn{1}{c|}{300} & \multicolumn{1}{c|}{\multirow{-3}{*}{192.7/14.7/25.1}} & \multirow{-3}{*}{4607/179/444} \\ \hline
\multicolumn{1}{l|}{{\color[HTML]{036400} }} & \multicolumn{1}{c|}{} & \multicolumn{1}{c|}{} & \multicolumn{5}{c|}{} & \multicolumn{1}{c|}{} & \multicolumn{1}{c|}{} & \multicolumn{1}{c|}{} &  \\
\multicolumn{1}{l|}{\multirow{-2}{*}{{\color[HTML]{036400} \begin{tabular}[c]{@{}l@{}}NTRUEncrypt$^*$\\ \cite{5418649}\end{tabular}}}} & \multicolumn{1}{c|}{\multirow{-2}{*}{Virtex-E}} & \multicolumn{1}{c|}{\multirow{-2}{*}{}} & \multicolumn{5}{c|}{\multirow{-2}{*}{27292/5160/14352/-/-}} & \multicolumn{1}{c|}{\multirow{-2}{*}{21175}} & \multicolumn{1}{c|}{\multirow{-2}{*}{62}} & \multicolumn{1}{c|}{\multirow{-2}{*}{-/1.54/1.41}} & \multirow{-2}{*}{-/33/30} \\ \hline
\multicolumn{1}{l|}{{\color[HTML]{036400} }} & \multicolumn{1}{c|}{} & \multicolumn{1}{c|}{\small{client}} & \multicolumn{5}{c|}{} & \multicolumn{1}{c|}{} & \multicolumn{1}{c|}{443} & \multicolumn{1}{c|}{} &  \\
\multicolumn{1}{l|}{\multirow{-2}{*}{{\color[HTML]{036400} \begin{tabular}[c]{@{}l@{}}InvRBLWE$^*$\\ \cite{8660431}\end{tabular}}}} & \multicolumn{1}{c|}{\multirow{-2}{*}{Virtex-7}} & \multicolumn{1}{c|}{\small{server}} & \multicolumn{5}{c|}{\multirow{-2}{*}{5000/5000/1292/0/0}} & \multicolumn{1}{c|}{\multirow{-2}{*}{2542}} & \multicolumn{1}{c|}{455} & \multicolumn{1}{c|}{\multirow{-2}{*}{0.95/1.97/0.95}} & \multirow{-2}{*}{2.4/5/2.4} \\ \hline
\multicolumn{1}{l|}{{\color[HTML]{036400} RLWE$^*$\cite{DBLP:conf/ches/RoyVMCV14}}} & \multicolumn{1}{c|}{Virtex-6} & \multicolumn{1}{c|}{} & \multicolumn{5}{c|}{1536/953/-/1.5/1} & \multicolumn{1}{c|}{778} & \multicolumn{1}{c|}{278} & \multicolumn{1}{c|}{-/47.9/21} & -/37/16 \\ \hline
\multicolumn{1}{l|}{{\color[HTML]{036400} RLWE$^*$\cite{DBLP:conf/sacrypt/PoppelmannG13}}} & \multicolumn{1}{c|}{Virtex-6} & \multicolumn{1}{c|}{} & \multicolumn{5}{c|}{5595/4760/1887/7/1} & \multicolumn{1}{c|}{4757} & \multicolumn{1}{c|}{251} & \multicolumn{1}{c|}{57.9/54.9/35.4} & 71/67/43 \\ \hline
\multicolumn{12}{l}{$^*$ This scheme is PKE not KEM and only provides CPA security. All other schemes are CCA secure.} \\
\multicolumn{12}{l}{$^\dagger$ These schemes provides NIST-level-3 security.} \\
\multicolumn{12}{l}{$^{**}$ ENS~\cite{Ayesha-k2-reduction} = Slice $+$ DSP$\times100$ $+$ BRAM$\times196$ $+$ LUT$/4$}
\end{tabular}
}
\end{table}


{We compare the implementation result of Rudraksh with the state-of-the-art hardware implementations of the notable candidates, including Kyber, in Tab.~\ref{tab:sota}. We include area, ENS, and time-area-product in terms of \texttt{ENS}$\times$\texttt{execution time} (T$\times$A) after implementing it in both Xilinx Virtex-7 and Artix-7 FPGA.
The implementation results of Rudraksh on these two FPGAs are similar. In Artix-7, the required ENSs of Rudraksh's algorithms are slightly less than in Virtex-7, and the maximum frequency is less in Artix-7 compared to Virtex-7 as Artix-7 FPGA is optimized for low-area compact applications. 
Our KEM uses the least ENS compared to all other implementations of the schemes presented in Tab.~\ref{tab:sota}, except the implementation of CPA-secure PKE scheme RLWE proposed in~\cite{DBLP:conf/ches/RoyVMCV14}. Our KEM consumes 19$\times$, 2.3$\times$, and 4.3$\times$ less ENS than NTRU-based PKE scheme NTRUEncrypt, Ring-LWE-based lightweight PKE scheme InvRBLWE~\cite{8660431}, and RLWE~\cite{DBLP:conf/sacrypt/PoppelmannG13}, respectively, while providing CCA security.} 

{Rudraksh uses 4.3$\times$ less ENS and 1.9$\times$/1.7$\times$/2.2$\times$ less T$\times$A for KG/Enc/Dec compared to {a} compact version of Kyber~\cite{compact_kyber}. A recent work~\cite{HE2024167} proposes an area-optimized implementation of Kyber at the cost of latency. Our design utilizes 2.9$\times$ less ENS and 14.8$\times$/16.6$\times$/19.6$\times$ less T$\times$A for KG/Enc/Dec with respect to Kyber~\cite{HE2024167}. Rudraksh needs 7.3$\times$ less ENS, the T$\times$A for KG/Enc/Dec are 1.6$\times$/1.8$\times$/2.1$\times$ less than the speed-optimized implementation of Kyber~\cite{9926344}.  A RISC-V-based softcore is used for Kyber implementation in~\cite{DBLP:journals/iacr/BanerjeeUC19}. It offers flexibility and re-usability but has a significantly high area and latency overhead. Rudraksh requires 10.6$\times$ less ENS and 584$\times$/805$\times$/744$\times$ less T$\times$A for KG/Enc/Dec compared to Kyber implementation in~\cite{DBLP:journals/iacr/BanerjeeUC19}.} 


{Compared to Frodo~\cite{frodo_hardware}, Rudraksh requires 5$\times$ less ENS, and 1829$\times$/1248$\times$/1288$\times$ less T$\times$A for KG/Enc/Dec. Rudraksh uses 2.9$\times$ less ENS than the efficient hardware implementation of NewHope~\cite{intt_by_2}. Although the T$\times$A for key-generation is approximately the same for Rudraksh and NewHope~\cite{intt_by_2}, it is 1.4$\times$ less in Rudraksh for encapsulation, and it is 2.1$\times$ more in Rudraksh for decapsulation. NewHope is an RLWE-based KEM, so its decapsulation algorithm performs fewer operations than Rudraksh (3, 512-length polynomial multiplication in NewHope and  99, 64-length polynomial multiplication in Rudraksh). So, the decapsulation operation of NewHope is faster than Rudraksh, and it impacts T$\times$A. However, NIST has shown {preference} while selecting Kyber against NewHope, mentioning that the RLWE-based scheme is most structured compared to any MLWE-based scheme, which is intermediately structured and closer to the standard-LWE~\cite{nist_3rd_round}. Therefore, Kyber is at least as secure as NewHope.}
{With respect to LightSaber~\cite{DBLP:journals/tches/RoyB20}, Rudraksh uses 5.7$\times$ less ENS and 1.9$\times$/2.2$\times$/2.4$\times$ less T$\times$A for KG/Enc/Dec. Very recently, full hardware implementation results of MLWR-based schemes Espada and Sable, RLWR-based scheme are presented in~\cite{scabbard-tecs}. However, these implementation results are only available for security version NIST-level-3. Therefore, we use them for comparison. Compared to Espada, which also uses $64$ length polynomials, Rudraksh requires 11$\times$ less ENS and 18.8$\times$/24.5$\times$/29.8$\times$ less T$\times$A for KG/Enc/Dec. With respect to Sable, Rudraksh uses 4.2$\times$ less ENS and 1.5$\times$/1.4$\times$/1.5$\times$ less T$\times$A for KG/Enc/Dec. Rudraksh requires 19.3$\times$ less ENS and 10.1$\times$/15.8$\times$/20.1$\times$ less T$\times$A for KG/Enc/Dec compared to Florete.}

{Further, we compare our implementation of Rudraksh with some NTRU-based KEMs, which provide NIST-level-1 security. Compared to NTRU-HRSS701~\cite{DBLP:journals/tc/DangMG23}, Rudraksh requires 23.9$\times$ less ENS and 76.6$\times$/1.4$\times$/7.3$\times$ less T$\times$A for KG/Enc/Dec. Rudraksh uses 21.5$\times$ less ENS and 76.8$\times$/2.3$\times$/4.9$\times$ less T$\times$A for KG/Enc/Dec with respect to NTRU-HPS677~\cite{DBLP:journals/tc/DangMG23}.}
%
%
{In brief, although the hardware implementations of our proposed lightweight CCA-secure quantum-secure design Rudraksh mainly focus on optimizing resources, it provides comparable speed and time-area products with respect to the implementations of state-of-the-art schemes. This makes Rudraksh very suitable for resource-constraint edge devices.}

\section{Conclusion and future work}\label{sec:conclusion}
{In this work, we performed a hardware-driven design space exploration based on resource consumption and proposed a design of MLWE-based KEM Rudraksh at the global minima of hardware requirement. Our design strategy involves optimizing the scheme's parameters and other design elements with continuous feedback from the implementation -- the final design results from multiple iterations and refinement of this process. The use of ASCON in Rudraksh as a lightweight XOF is also the first of its kind.} 
{Although ASCON is a lightweight standard for hash and XOF, its small state size affects the overall efficiency. This work solves this problem with a low critical path design to achieve very high frequency. It also consumes low power thanks to simpler circuits and lower state size. Finally, we synchronize the cycles of ASCON-XOF and operate in parallel with NTT to reduce the overall latency.} 

Our immediate next plan is to design an ASIC of our PQ KEM and compare the results. {It is also interesting to compare the implementation cost of Rudraksh on small microcontrollers, such as Cortex-M0, Cortex-M4, etc. There are enormous possibilities in developing a lightweight lattice-based KEM. For example, this work is limited to designing a KEM based on the (R/M)LWE hard problem; a similar strategy can also be applied to the KEM based on {the} NTRU hard problem. The same approaches can be used to design a lightweight lattice-based digital signature scheme. We plan to work on these topics in the future.}


{We would like to mention that we observed that LWE-based KEMs benefited us more than LWR-based KEMs in designing lightweight cryptography with our implementation method. Therefore, we explored the design space of LWE-based KEM. However, the possibilities of designing optimized KEMs are limitless. For example, some KEMs introduced recently, such as SMAUG~\cite{smaug_kem} and TiGER~\cite{TiGER} use a combination of (R/M)LWE and (R/M)LWR problems. Different explorations of design spaces can provide better designs in different aspects. The benefit of using ASCON over Keccak in the above-mentioned schemes and different LWR-based KEMs also needs to be explored. Exploring all these possibilities is very difficult to cover in a single work. Therefore, we would like to emphasize the importance of more research in this direction.}    


On another note, side-channel attack (SCA) protection is necessary for widely deployed algorithms. The implementation of Rudraksh is constant-time. Therefore, it is already timing {SCA} secure. One widely used provably secure countermeasure against other {SCAs} is masking. We need some additional components for a masked version of Rudraksh, namely masked ASCON, masked CBD, arithmetic-to-Boolean (A2B), and Boolean-to-arithmetic (B2A) conversion algorithms~\cite{FO-masked-kyber-HeinzKLPSS22}. ASCON is more side-channel resilient than other lightweight schemes, and the area overhead of SCA-protected ASCON with masking will be comparatively {lower} than Keccak~\cite{ASCON}. It will benefit our scheme by reducing the area cost of side-channel protection with masking. The cost of the area consumption of masked CBD, A2B, and B2A will be approximately the same for Rudraksh and Kyber. Therefore, the overall area consumption of {the} masked Rudraksh should be lower than {that of} Kyber. {As masked (R/M)LWR-based schemes perform better than (R/M)LWE-based schemes thanks to power-of-2 moduli{, it} is interesting to compare the implementation cost of masked Rudraksh with masked Saber~\cite{FO-masked-saber, HO_mask_Saber} or masked Scabbard~\cite{masked-scabbard}}. However, it needs {more} formal and experimental verification, which we have left {for} future work. 
{Furthermore, recently proposed circuit-level techniques~\cite{sen2024circuit}, such as signature attenuation~\cite{ghosh202136, ghosh2021syn, ghosh2022digital} and clocking methods~\cite{ghosh2023power,ghosh2024exploiting}, can serve as effective countermeasures, often offering lower overhead than standard masking techniques. These approaches need further exploration in future research, particularly in enhancing side-channel security for lightweight cryptographic schemes.}

\noindent\textbf{Acknowledgements.} This work was partially supported by Horizon 2020 ERC Advanced Grant (101020005 Belfort), Horizon Europe (101070008 ORSHIN), CyberSecurity Research Flanders with reference number VOEWICS02, BE QCI: Belgian-QCI (3E230370) (see beqci.eu), and Intel Corporation. The work of Angshuman Karmakar is supported by the Research-I foundation from Infosys, the Initiation grant from IIT Kanpur, and the Google India research fellowship. The work of Archisman Ghosh is supported by the NSF (Grant CNS 17-19235), TSMC Center for Secure Microelectronics Ecosystem (CSME) and Intel Corporation.  

We thank Jonas Bertels for the interesting discussions regarding NTT designs.

\bibliographystyle{alpha}
\bibliography{main}

\end{document}